\documentclass{article}

\usepackage{arxiv}

\usepackage[utf8]{inputenc}
\usepackage[T1]{fontenc}
\usepackage{url}
\usepackage{booktabs}
\usepackage{amsfonts}
\usepackage{nicefrac}
\usepackage{float}

\usepackage{microtype}
\usepackage{graphicx}
\usepackage{graphics}
\usepackage{caption}
\usepackage[super]{nth}
\usepackage{tabularx}
\usepackage{pdflscape} 
\usepackage{hyperref}
\hypersetup{colorlinks,citecolor=blue,urlcolor=blue}

\usepackage[hang,flushmargin]{footmisc}

\usepackage{natbib}
\newdimen\origiwspc
\newdimen\origiwstr

\title{{\em Pandemic News:} \\Facebook Pages of Mainstream News Media and the Coronavirus Crisis -- A Computational Content Analysis}

\author{
 Thorsten Quandt* \\
  Department of Communication\\
  University of Münster\\
  Münster, 48143 Germany \\
  \texttt{thorsten.quandt@uni-muenster.de} \\
  \And
  Svenja~Boberg \\
  Department of Communication\\
  University of Münster\\
  Münster, 48143 Germany \\
  \texttt{svenja.boberg@uni-muenster.de} \\
 \And
 Tim Schatto-Eckrodt \\
  Department of Communication\\
  University of Münster\\
  Münster, 48143 Germany \\
  \texttt{tim.schatto-eckrodt@uni-muenster.de} \\
  \And
  Lena Frischlich \\
  Department of Media and Communication\\
  LMU Munich\\
  Munich, 80538 Germany \\
  \texttt{lena.frischlich@uni-muenster.de} \\
}

\begin{document}
\maketitle

\begin{abstract}
The unfolding of the COVID-19 pandemic has been an unprecedented challenge for news media around the globe. While journalism is meant to process yet unknown events by design, the dynamically evolving situation affected all aspects of life in such profound ways that even the routines of crisis reporting seemed to be insufficient. Critics noted tendencies to horse-race reporting and uncritical coverage, with journalism being too close to official statements and too affirmative of political decisions. However, empirical data on the performance of journalistic news media during the crisis has been lacking thus far. Therefore, the current study analyzes the Facebook messages of journalistic news media during the early Coronavirus crisis, based on a large German data set from January to the second half of March 2020. Using computational content analysis methods, reach and interactions, topical structure, relevant actors, negativity of messages, as well as the coverage of fabricated news and conspiracy theories were examined. 

The analysis reveals a nuanced picture. The topical structure of the near-time Facebook coverage changed during various stages of the crisis, with just partial support for the claims of critics. Indeed, the initial stages were somewhat lacking in topical breadth, but later stages offered a broad range of coverage on Corona-related issues and societal concerns. While political elites played a central role in these messages, there was also a wide spectrum of other actors present. Further, journalistic media covered fake news and conspiracy theories during the crisis, but they consistently contextualized them as what they were and debunked the false claims circulating in public. While some criticism regarding the performance of journalism during the crisis received mild empirical support, the analysis did not find overwhelming signs of systemic dysfunctionalities. Overall, journalistic media did not default to a uniform reaction nor to sprawling, information-poor pandemic news, but they responded with a multi-perspective coverage of the crisis.
\end{abstract}


\section{Introduction: Covering an unfolding crisis}\label{introduction}

News media frequently cover crisis situations, from military conflicts, natural disasters, and terrorist attacks to large-scale effects of man-made accidents. Some of these are unexpected and practically without a build-up (like the Chernobyl disaster or 9/11), but some of them evolve (like hurricane seasons), and some fall somewhere in-between (like conflicts between nations that may, at some point, lead to sudden military actions). For most types of crises \citep[][p. 89 ff]{Nord_Stroembaeck_2006}, journalism has developed specific routines and practices \citep[e.g.][]{Olsson_Nord_2015,Usher_2009}. Preparation and rules, as well as rituals \citep{Durham_2008,Riegert_Olsson_2007} help in dealing with unexpected events or unfolding situations that fall outside the coverage of “normal” times. Part of this is driven by informational goals that are inherent to journalism, but part of it is an urge to establish authority in societal communication under non-normal conditions \citep{Durham_2008,Riegert_Olsson_2007}. 

Online channels have further contributed to this due to directly observable, competing information by journalists and other sources in a highly complex news environment \citep{Boczkowski_2010,Lim_2013}. (Near-)Real-time (online) coverage has become the norm in times of the 24-hours news cycle \citep{Rosenberg_Feldman_2008,Saltzis_2012}, and this has propelled the need of processing any type of event as it happens and the development of rules to reliably deal with the unexpected immediately \citep{Buhl_Guenther_Quandt_2019}. Indeed, many journalistic media provide “breaking news” and other information not only via their main sites, but also via social media channels (like Facebook or Twitter), directly and instantaneously addressing users in that environment and thus contributing to a unique and distinct “information space” \citep{Schmidt_Zollo_Vicario_M._A._A._Stanley_E_Quattrociocchi_2017}. One would expect such a highly developed and complex system to be ready to tackle any type of occurrence.

The Corona crisis, however, seems to have caught many news organizations off-guard; the coverage during the initial spread of the virus in early 2020 was scarce \citep[see also ][]{Boberg_Quandt_Schatto-Eckrodt_Frischlich_2020}, and many news media around the globe made the false assessment that the outbreak would either not affect countries other than China or that the virus was less dangerous than the flu \citep[e.g.,][]{Henry_Hauck_2020}. Coverage did not take off until the consequences for the respective nations became apparent and experiential. When the virus and its effects reached the national populations, though, the news output on the virus threat seemed to explode \citep{Boberg_Quandt_Schatto-Eckrodt_Frischlich_2020}, and journalism switched to full “crisis mode.”

Critics, including journalism researchers, noted multiple issues with this reaction, identifying deficits in the coverage and journalism’s functioning. There were national differences in criticism, though, ranging from “too late,” “too much,” or “too narrow” to “not enough plurality” or “heroization of singular actors.” In the case of Germany, which is the country of analysis of this paper, communication scholars noted a limited set of expert voices, “system journalism” and “court circular”\footnote{In German, “Hofberichterstattung” (court circular) means news of the court and the royal family, but it is also used (in a figurative sense) to denote uncritical reporting on the political elite and their decisions.}  for some media \citep{Jarren_2020}, a lack of distance and critique, not enough variance, “horse-race reporting” with regard to national comparisons of infection numbers \citep{Meier_Wyss_2020}, the use of war or competitive rhetoric \citep{Brost_Poerksen_2020,Jarren_2020,Meier_Wyss_2020}, as well as “gasping” journalism and a language of “no alternatives” \citep{Brost_Poerksen_2020}. Inadvertently, some of these critical voices received “applause from the wrong side” \citep{Russ-Mohl_2020}, including left and right extremists, as they fit a system-critical narrative, which is also present in alternative news media \citep{Boberg_Quandt_Schatto-Eckrodt_Frischlich_2020,Holt_Ustad_Figenschou_Frischlich_2019}. In turn, journalists responded to the criticism as well, calling it “unfortunate” and “desperately finding a fly in the ointment” \citep{Zoerner_2020} or “going nowhere” and partially “grotesque” \citep{DInka_2020}. 

Indeed, there has been a lack of empirical research supporting such early criticism, as these initial reactions were based on individual observations and opinions. However, the assumption of extreme events challenging journalism and its function is not unfounded. Previous research has shown that there are “frame breakers” \citep{Lund_Olsson_2016}, that is, the “worst events [...] that are genuinely unprecedented and shocking,” which cannot be “anticipated, and challenge even the most established journalistic practices” \citep[][p. 358]{Lund_Olsson_2016} and in essence, lead to inadequate responses. In previous crisis cases, empirical work has shown that this fosters cases of “pseudo-journalism” \citep[][p. 105 f.]{Nord_Stroembaeck_2006} and a “reporting more, informing less” situation \citep[][p. 85]{Nord_Stroembaeck_2006}. 

The anecdotal evidence of individual observations and the assumptions from previous research have yet to be tested against empirical research, which sets the goal for the current working paper. Here, we will analyze the posts of journalistic media on Facebook in the (early) Corona crisis (specifically, in Germany). The choice of Facebook news is deliberate and two-fold; by doing so, we can give a virtually complete overview of the near-time news of the most relevant journalistic news media during the early crisis (see also Section \ref{methods}). Facebook is a highly relevant distribution channel for news media, as it is still ahead of other message-based social media platforms in Germany, with 21\% of the population above the age of 14 using it daily \citep[in contrast to highly researched Twitter, which is only used by 2\%;][]{Beisch_Koch_Schaefer_2019}. Further, it allows for a comparison with our earlier working paper \citep{Boberg_Quandt_Schatto-Eckrodt_Frischlich_2020} that focused on the message output of “alternative news media” who position themselves as a “corrective of traditional,” “legacy,” or “mainstream news media” \citep{Holt_Ustad_Figenschou_Frischlich_2019}. This is particularly interesting, as the alternative news media have been critical of the traditional mainstream media’s performance during the crisis and claimed to give a more truthful depiction of the events and to offer a broader choice of voices, including oppositional ones that questioned public statements and the actions of the government and officials \citep[for an in-depth analysis, see][]{Boberg_Quandt_Schatto-Eckrodt_Frischlich_2020}. The current study allows a closer inspection of such claims, as well as the above-mentioned public criticism by media scholars, and a more substantive basis for the analysis of the performance of journalism during the (early) crisis.

This working paper is deliberately limited to the case of Germany and the initial weeks of the pandemic (until March 22, which was the date of decision to introduce a national prohibition of contact and other measures to “flatten the curve” of infections); this is exactly the same sample period as in our previous working paper on alternative news media. Both papers are part of an ongoing research effort that will expand the analysis period and focus in subsequent publications. We also hope to add to the international research efforts on news and communication in the Corona crisis, answering more general questions on the performance of news media in the crisis. In particular, following previous research on journalist’s crisis reporting \citep{Nord_Stroembaeck_2006}, we will ask whether journalistic media contributed to a parallel “infodemic” \citep{WHO_2020b,Zarocostas_2020} in the form of sprawling, but information-poor, \emph{pandemic news}.

\section{Crisis coverage and the transformation of the (news) media system}\label{crisis_coverage}

\subsection{News media in times of crisis}\label{news_media_in_crisis}

Crisis coverage has been subject to journalism research for decades, often in relation to specific events. In particular, in the aftermath of the 9/11 attacks and other terrorist events and the military conflicts that directly or indirectly followed it, there has been a rising interest in the analysis of the journalistic response and performance throughout extraordinary times \citep{Lund_Olsson_2016}. While a major block of this research focused on armed conflict and shocking events tied to terrorism, some of the findings may be helpful in understanding the Corona crisis coverage. Naturally, differing qualities of the respective cases of application still need to be kept in mind, in particular, with regard to time-based developments and the existence of tangible conflict groups and interests (or the lack thereof). Indeed, there are limitations to applying the patterns as observed in armed conflicts and attacks to the pandemic situation. German virologist and government advisor Christian Drosten called the pandemic a “natural disaster in slow-motion” \citep{NDR_Info_2020}, a statement that was later on criticized by other experts. However, in terms of news related properties, this characterization is quite fitting. Many other “frame breakers” are structured around “events” that happen at specific points of time. The time span of the relevant trigger events is often minimal, without a lengthy build-up, and the effects are directly apparent and severe (like in the case of terror or military attacks, volcano eruptions, or tsunamis). The pandemic, on the other hand, evolved over a long period of time, as Drosten pointed out, with no observable and discrete trigger event (i.e., the first infection cannot be observed and displayed by media, and just becomes a relevant “invisible” event in hindsight), and there are no immediately observable, direct, and strong effects following the initial event. In that sense, the pandemic reminds us more of the North American hurricane season than, for example, a military conflict. However, the hurricane season is a recurring and highly predictable natural catastrophe, so the media has time to prepare for what is, in essence, an expectable disaster \citep[while the individual hurricane still can be much more catastrophic than previewed; for analyses of the coverage of Hurricane Katrina, see][]{Robinson_2009, Usher_2009}. The pandemic, however, took societies and journalism around the world off guard, and as such, it was unprecedented in its effects, not only on the health of individuals, but also on society at large.    

As noted above, other frame breakers were discussed in journalism research \citep{Boin_t_Hart_Stern_Sundelius_2005,Lund_Olsson_2016,Olsson_2010}, following \citeauthor{Weick_1993}'s \citeyearpar{Weick_1993} characterization of “cosmology episodes” that shatter the universe’s order (in a figurative sense) and also the means to piece that order together again. Journalism research has identified and analyzed multiple potential frame breakers, like 9/11 and other terrorist attacks, and wars, like the one in Afghanistan or Iraq \citep{Lund_Olsson_2016, Nord_Stroembaeck_2006}. Many of these case studies scrutinize the effect of preparations, the importance of routines and standards, as well as the role of time in the unfolding of such events. Indeed, multiple studies show that norms, standards and roles in journalism, as well as established work routines, help in channeling events that are unknown and untypified when they enter journalistic processing in crisis situations \citep[e.g.,][]{Olsson_2009, Olsson_Nord_2015, Usher_2009}. Expecting the unexpected is, in many ways, part of the modus operandi of journalism even under normal conditions \citep[i.e., as routinized, rule-based news processing; see][]{Ryfe_2006,Tuchman_1973}, but as research shows, even for extraordinary events, the level of disruption can be reduced by having appropriate work routines in place. 

Other studies have focused on rituals in journalism \citep{Durham_2008, Riegert_Olsson_2007}, which is different from a functional viewpoint that is primarily interested in the information processing side. It has been argued that rituals are an essential aspect of journalism’s functioning in crisis \citep{Couldry_2003,Durham_2008}, as “media rituals can be said to function as a way for media organizations (...) to establish authority by playing a key part in society’s healing process” \citep[][p. 147]{Riegert_Olsson_2007}.  Based on the analysis of broadcast journalist’s decision making during two crises (9/11 and the Anna Lindh murder in 2003), \citet[][p. 143]{Riegert_Olsson_2007} conclude that “in times of crisis, the roles of psychologist, comforter and co-mourner should be considered journalistic role conceptions especially in a live, 24-hour news culture.” While being predominantly interested in the symbolic function of journalism, they still stress the importance of specific journalistic strategies of dealing with the unknown under extreme time pressure.

\citet{Nord_Stroembaeck_2006}, in their analysis of Swedish news media in multiple crises, noted that time is one key component in the way media operate in crises and that it can affect the quality considerably. Their study hypothesized that the performance and quality of journalistic reporting in crisis is “dependent on both the existence of media routines and the possibility for the media to make adequate preparations” \citep[][p. 89]{Nord_Stroembaeck_2006}, further typifying crises along these two dimensions (media preparations and media routines). However, they found that even with a long build-up time and the existence of proper routines, the quality of the coverage can suffer: “The larger the news hole devoted to covering an event and the more the different media report about an event, the higher the risk that striving for correct news will be less important than striving for fast and attention-grabbing news, and the higher the risk that journalism will be replaced by pseudo-journalism” \citep[][p. 107]{Nord_Stroembaeck_2006}. So, rather than “time” alone, the news output in a given time span may be a more decisive factor, which seems logical, given a limited amount of (human) resources in journalistic media. Taking the expansion of online information channels in recent years and the corresponding changes in and adaption of journalistic news into account (see Section \ref{new_platforms}), the pressure to “fill empty space” under time constraints has plausibly even increased, as have the resulting dysfunctionalities.  
 
\subsection{New platforms, the evolution of journalism, and the compression of time}\label{new_platforms}

The efforts to report on extraordinary events in the shortest possible time predate the Internet by far, though, and to be as up-to-date as possible has always been relevant for news production \citep{Galtung_Ruge_1965, Lewis_Cushion_2009}. However, newspapers were bound to deadlines, due to fixed production and publication rhythms, and linear broadcasting was also following specific schedules. Non-routine and spectacular cases of “what-a-story” \citep{Berkowitz_1992}, like the Kennedy assassination \citep{Greenberg_1964, Rosenberg_Feldman_2008}, were exceptions to the rule. Here, the broadcasters stopped their normal schedule for special programs that offered immediate news as it broke. Such news events were not the norm, though \citep{Rogers_2000}. The arrival of 24-hour TV news channels changed this considerably, as the threshold for the immediate coverage of “breaking news” was lowered systematically \citep{Lewis_Cushion_2009, Rosenberg_Feldman_2008}. 

The Internet and online news media further contributed to the evolution of journalism—the “online first” dissemination of news in a continuous process ushered in a new “era of fluid deadlines” \citep{Buhl_Guenther_Quandt_2019}. Such fluidity not only implies immediacy as a central norm \citep{Lim_2012, Quandt_Loeffelholz_Weaver_Hanitzsch_Altmeppen_2006}, it is also a potentially continuous updating and editing of news \citep{Saltzis_2012}, the constant observation of competitors \citep{Boczkowski_2010, Lim_2013} in the online news ecosystem \citep{Anderson_2010, Buhl_Guenther_Quandt_2018}, and a shortening of news cycles \citep{Rosenberg_Feldman_2008} down to a minimum. New platforms like Facebook and Twitter further added to this, as they are built for constant information flows; the timeline logic, extremely short messages, a culture of pushing texts in an evolving stream as you think or write them have all contributed to a further compression of time and journalistic processes, and even a vanishing of deadlines \citep{Karlsson_Stroembaeck_2010}, not just for extraordinary events. In such an environment, news needs to be published quickly and under high pressure—even more so if there is a demand for a high volume of news on a specific topic and if competing information providers virtually react in parallel (as is the case in the pandemic). 

Journalism research has focused on the effects of the news diffusion processes, and by using computational methods, very detailed tracking and analysis of the spread of news over time has become possible. For example, \citet{Harder_Sevenans_Van_Aelst_2017} applied a “news story” approach to the analysis of intermedia agenda-setting processes in hybrid media systems. \citet{Buhl_Guenther_Quandt_2019, Buhl_Guenther_Quandt_2018} showed that specific event features (like negativity and prominence) lead to extremely fast reaction times and stories by a high number of media. They labeled such “explosive” ecosystem reactions as “news bursts.” News diffusion in online media has also been described as “cascades” \citep{Lotan_Graeff_Ananny_Gaffney_Pearce_others_2011} or “news waves” \citep{Waldherr_2014}. This empirical research not only stresses the effects of online media and platforms on virtually immediate news diffusion processes, but also the complexities of information distribution in a still evolving hybrid media system \citep{Chadwick_2017}, where older and newer media logics are intertwined. Further, it reveals that information in such a system seems to partially spread in a viral, uncontrolled logic;  at the same time, event features are relevant, and actors (like news media) still play an active, partially steering role. 

In that sense, the discussion of an “infodemic” in relation to the Corona crisis is only partially specific to the pandemic; it’s also based on changes in the media system and journalism that are ongoing for much longer. Therefore, the current research is not only informed by the current debate and criticism of journalism in the crisis, but it can be also linked to the ongoing research efforts in understanding evolving information flows in a complex media system. 
 
\section{Research questions}\label{research_questions}

As discussed in the previous sections, we are focusing on traditional “mainstream” news and their Corona-related Facebook activities during the early Corona crisis in Germany, based on an existing body of research on crisis journalism and the evolving media system. This research has shown that under certain circumstances, journalism switches to forms of crisis coverage that may be insufficient for the demands of extraordinary events and even countering the goals of journalism in open, democratic societies. It has been argued that these dysfunctionalities actually stem from seemingly “logical” reactions and some journalistic rules and routines that may be fully functional in a “normal” setting (see Section \ref{crisis_coverage}). Indeed, there has been criticism of journalism’s Corona coverage, noting a lack of initial reactions, and then a sudden growth of uncritical “system journalism,” horse-race reporting, a lack of distance from politics, decision makers, and experts, with a focus on a very selective number of actors, which were uncritically heroized as the ones “battling the virus,” and a very limited range of topics and viewpoints—in essence. restricting pluralism and the voices to be heard in public (see Section \ref{introduction}). However, empirical evidence to bolster this criticism is still lacking. Therefore, we will apply a computational content analysis to have a closer look at the message structure of journalistic news media in the Corona crisis. In line with our previous research on alternative news media \citep{Boberg_Quandt_Schatto-Eckrodt_Frischlich_2020}, we will ask a few essential base questions:

\hspace{15pt}[RQ 1a]\quad How broad is the reach of journalistic news media’s Corona coverage on Facebook?

\hspace{15pt}[RQ 1b]\quad How many interactions do they evoke?

\hspace{15pt}[RQ 2a]\quad What are the central topics of their coverage of the Corona crisis?

\hspace{15pt}[RQ 2b]\quad How did the topic structure evolve over time?

\hspace{15pt}[RQ 3a]\quad Who are the most prominent actors in their coverage?

\hspace{15pt}[RQ 3b]\quad In what topical contexts are these actors mentioned in the coverage?

In addition to these RQs, which parallel the analyses in our previous paper, there is also the open question of whether the reporting was predominantly affirmative of the political and, more generally speaking, “systemic” reactions, picking up the public debate on journalism’s performance as discussed in Section \ref{introduction}. Such a stance can be expressed through language referring to affective states and emotions in the given topical contexts and in relation to specific actors, and in particular, by a lack of critical, negative evaluations of said entities and their actions. On a very broad level, this can be analyzed through sentiment analysis; therefore, we also ask:  

\hspace{15pt}[RQ 4]\quad What are the sentiments expressed in the Corona reporting?  

Finally, there has been public criticism regarding the truthfulness of the Corona coverage in mainstream journalism. This has been especially prominent in the sphere of alternative news media, who claimed officials and “system media” lied to the people. Conversely, alternative news media themselves were also criticized for resorting to fake news and conspiracy theories. Indeed, there is partial empirical support for them spreading misleading information and some conspiracy theories during the early Corona crisis \citep[see][]{Boberg_Quandt_Schatto-Eckrodt_Frischlich_2020}. Therefore, we will check whether the same pattern can be found for traditional mainstream media, as claimed by critics, using the exact same analysis methods for the sake of comparability and adding the following research question:

\hspace{15pt}\hangindent=15pt[RQ 5]\quad Do the messages of journalistic news media include fake news or conspiracy theories, as identified by fact-checking entities? 

These questions are certainly not exhaustive, and they are kept simple for the purpose of a working paper portraying news media activities on Facebook during the first weeks of the crisis. We also deliberately refrained from intricate hypothesis testing, which we reserved for more detailed follow-up papers. However, for the time being, the questions should allow for an initial insight into the performance of news media during the given time period, and by answering them, we should get a clearer impression of whether journalistic media were spreading a large amount of “thin” pandemic news or whether they offered a rich and plural mix of information.   

\section{Methods}\label{methods}

\subsection{Sample and data}\label{sample_and_data}

In the subsequent analysis, we address the Facebook posts of news outlets during the early Corona crisis beginning with January 7, 2020, the day Chinese authorities first isolated the new type of virus \citep{WHO_2020a}, until March 22, 2020. We relied on the same data collection as in the previous study on alternative news media \citep{Boberg_Quandt_Schatto-Eckrodt_Frischlich_2020}, now focusing on mainstream news media. As part of Facebook's transparency initiative, we were able to collect the data via CrowdTangle, a platform that tracks the public (non-targeted, non-gated) content from influential or verified Facebook profiles and pages (information on private user content is not provided). 

For the collection of mainstream media posts, we relied on an updated version of an earlier overview of German mainstream newspapers \citep{Frischlich_Boberg_Quandt_2019}. The procedure followed a multi-level triangulation approach identifying online media sites with more than 500,000 unique users \citep{AGOF_2020}, then detecting online newspapers within this database following the literature \citep{Buhl_Guenther_Quandt_2018, Schuetz_2012} and our own investigations, excluding all non-journalistic outlets \citep[for a similar approach, see][]{Steindl_Lauerer_Hanitzsch_2017, Weischenberg_Malik_Scholl_2006}. Hereby, we identified 80 national and regional online newspapers of which 78 ran one or more active Facebook pages during our analysis period. In order to avoid a skewed sample, we only included the main page of each news outlet, resulting in a total of 100,432 mainstream media posts. For purposes of comparison, the dataset also included all Facebook posts by German fact-checking sites, Mimikama and Correctiv (\emph{n} = 1,355) and 32 alternative media outlets (\emph{n} = 15,207) \citep[for a detailed description of the sample, see][]{Boberg_Quandt_Schatto-Eckrodt_Frischlich_2020}.

\subsection{Preprocessing}\label{preprocessing}

CrowdTangle allows for a typical set of metadata like timestamps, page likes, engagement metrics (number of likes, shares, and comments), and information on linked URLs in the posts (including embedded headlines and teaser texts of the linked sites or articles). In order to analyze the coverage of COVID-19, we extracted posts based on regular expressions including different spellings and technical terms for “Corona” and “COVID-19” and topic-related terms like “epidemic,” “pandemic,” and “quarantine” or often-used hashtags like “\#flattenthecurve,” “\#washyourhands,” and “\#stayhome.” Our final sample, including all posts covering Corona, consisted of 18,051 mainstream, 2,446 alternative news media, and 282 fact-checking posts. In order to determine reach, we relied on the number of page likes on the date of the data collection as provided by CrowdTangle. The number of total interactions were computed by the unweighted addition of number of likes, shares, comments, and reactions into a new variable. 

To make the data ready for analysis, we used a pipeline of preprocessing steps that were tried and tested in previous studies \citep{Boberg_Schatto-Eckrodt_Frischlich_Quandt_2018, Guenther_Quandt_2016}, which included removing fragments of html-markup, URLs, mentions, and hashtags, punctuation, and stopwords. The ambiguous use of words was disentangled by identifying named entities with the help of the Stanford CoreNLP Toolkit \citep{Manning_Surdeanu_Bauer_Finkel_Bethard_McClosky_2014} and adjusting them manually. This included the cleansing of different names (e.g., “the US-president” and “Donald Trump”), but also terms used in different contexts (e.g., “government,” “German government,” or “US-government”).

\subsection{Analytical approach}\label{analytical_approach}

The content analysis of mainstream news media Facebook posts followed the exact same analytical approach as demonstrated in our working paper on alternative news media \citep[for a detailed description, see][]{Boberg_Quandt_Schatto-Eckrodt_Frischlich_2020}, only adding a sentiment analysis in order to shed light on the use of affective language by journalists. 

To capture the thematic structure (RQ 2), we applied the inductive method of topic modelling. The algorithm of  Latent Dirichlet allocation \citep[LDA;][]{Blei_Ng_Jordan_2003} is able to recognize patterns in words that frequently occur together and to generate different topics from these patterns. This procedure is able to detect to what extent each word of the documents characterizes the respective topic (\emph{$\beta$}) and how far each topic is present in each document (\emph{$\gamma$}), so each document can be represented by a mixture of different topics \citep{Guenther_Domahidi_2017}. Since the number of topics has to be determined in advance, a number of models with varying \emph{k} (2--50) were calculated to find a number of topics that makes the most sense in terms of mathematical measures \citep[e.g., via the R ldatuning package][]{Nikita_Chaney_2016}, but also in terms of content and meaning. This resulted in a 12-topic solution, which was underlined by the inspection of different topic solutions with larger \emph{k} (e.g. \emph{k} = 23), since more finely resolved topics do not show new topical structures, but only finer subdivisions of the 12-topic solution. The 12 topics were characterized by looking at the top terms and most representative documents for each topic (see Table \ref{tab:topic_description}) and validated using word and topic intrusion tests \citep{Chang_Gerrish_Wang_Boyd-graber_Blei_2009}.

In order to inspect the most central actors (RQ 3) and the contexts in which they are discussed, a co-occurrence analysis was conducted \citep{Bordag_2008}, which counts the number of times two words occur in the same document. The pairwise count was computed on the whole sample and then filtered by the 20 most frequent named entities as annotated during named entity recognition (for the whole list see supplemental materials). The co-occurrence network was then visualized as a force-directed graph using the R-package igraph \citep{Csardi_Nepusz_others_2006}.

The sentiment analysis is a dictionary-based procedure to identify positive or negative attitudes expressed in a text \citep[for an overview, see][]{Silge_Robinson_2017}. In order to extract the sentiments in our data (RQ 4) we used SentiWS, a comprehensive validated sentiment dictionary in German \citep{Remus_Quasthoff_Heyer_2010}. SentiWS comes with a list of positive and negative words and their inflections and the respective score for weighting from -1 to 1. After a first application of the dictionary, we decided to focus on the negative sentiments because we were mostly interested in the prevalence of negativity; also, it became apparent that the positive dictionary needed further validation in the context of Corona since many words, like “positive,” “increase,” “massive,” etc. are not likely to hint at a positive sentiment in the realm of a pandemic. Looking at the negative sentiments, the same problems became apparent when considering the exact weighted sentiment scores, so we decided to focus on posts that included negative sentiments without paying attention to the number of sentiment-related words or the exact sentiment score. Note that this is an ongoing research project and we aim at developing more fine-grained measurements and, in the scope of COVID-19, more applicable dictionaries.  

In the previous study on alternative news media, we used the fact-checking sites Mimikama and Correctiv to identify Corona-related fake news and conspiracy theories, which resulted in four stories (for a detailed description, see Table \ref{tab:topic_description}). In a dictionary-based approach, we checked whether these stories were also present in the data of the current study (RQ5) using the same keyword lists that were developed in the previous study (for a detailed description see, supplemental materials). After the keyword-assisted detection of posts that referred to the respective fake news or conspiracy theories, the documents were analyzed manually to summarize them in subordinate styles of representation to see how these stories were covered. 

\section{Results}\label{results}

\subsection{Reach and total interactions}\label{reach_and_interactions}

The analyzed corpus of 18,051 Corona-related posts triggered millions of likes, interactions, and shares (RQ1a and b; see Table \ref{tab:media_types}). However, the output of messages and reactions on the pandemic differed significantly over the various phases of the analysis period. 

\begin{table}[h]
 \caption{Media types}
  \centering
  \begin{tabularx}{\textwidth}{lr rr rr rr rr}
    \toprule
     & & \multicolumn{2}{c}{Page likes} &
     \multicolumn{2}{c}{Posts} &
     \multicolumn{2}{c}{Interactions} & \multicolumn{2}{c}{Shares} \\
     \cmidrule(lr){3-4} \cmidrule(lr){5-6} \cmidrule(lr){7-8} \cmidrule(lr){9-10}
    Media Type & Pages & Total & Per page & Total & Per page & Total & Per post & Total & Per post \\
    \midrule
    Mainstream   & 78 & 16,779,317 & 215,119 & 18,051 & 231 & 5,732,155 & 318 & 1,472,648 & 82 \\
    Alternative  & 32 & 1,653,208 & 51,663 & 2,446 & 76 & 589,534 & 241 & 197,401 & 81 \\
    \bottomrule
 \end{tabularx}
 \label{tab:media_types}
\end{table}

\begin{figure}
  \centering
  \includegraphics[width=1.0\textwidth]{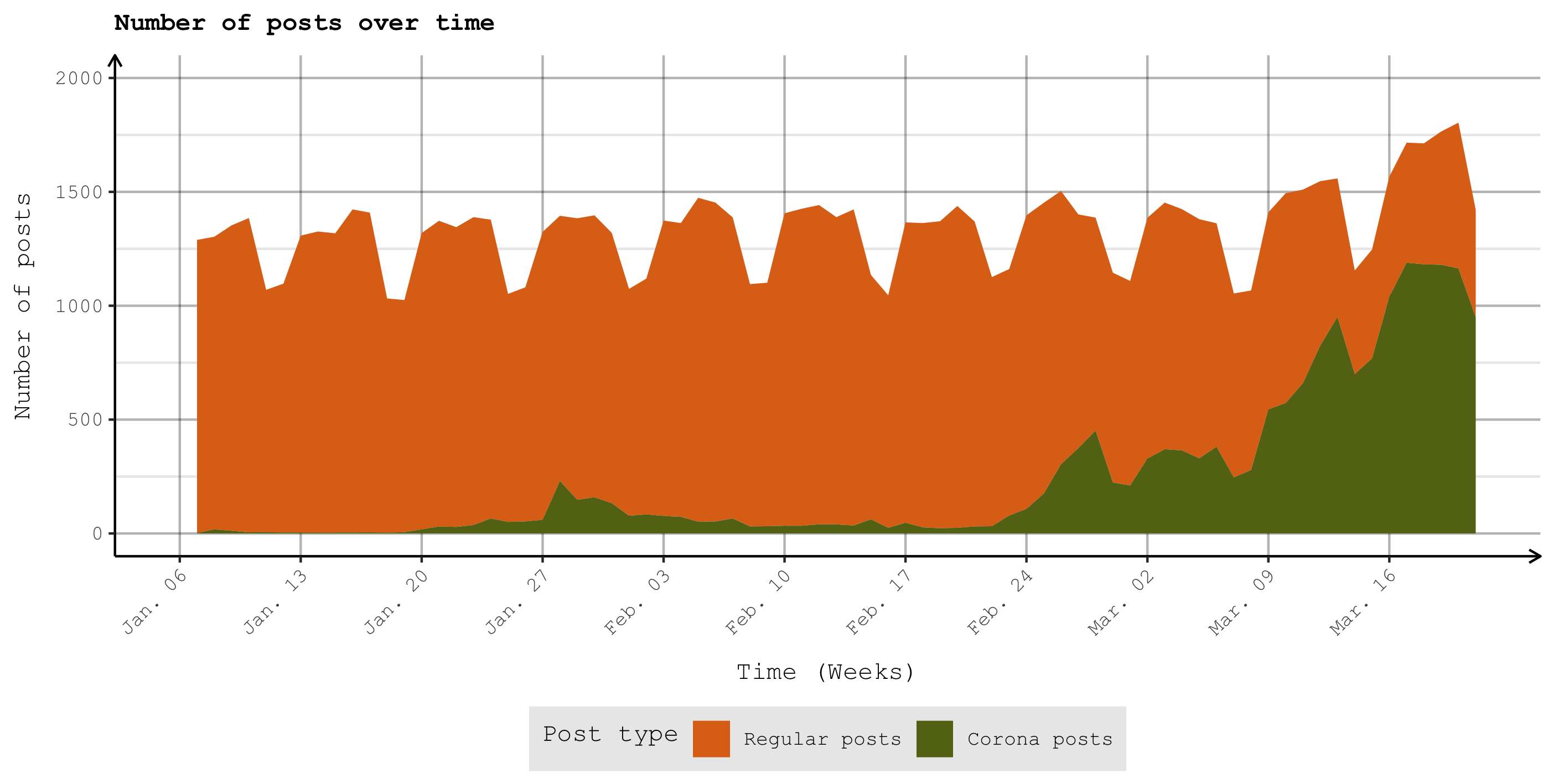}
  \caption{Number of posts over time}
  \label{fig:posts_over_time}
\end{figure}

A first time-based analysis of the overall message flow, including all the 100,432 messages of the 78 media under analysis, showed a relatively stable pattern and average volume of the overall news output (Figure \ref{fig:posts_over_time}). The recurring “dents” in the output immediately strike the eye. Such sudden drops can be explained by reduced (or no) staff in the newsrooms during the weekend and the respective weekend production routines, which are obviously still in place even in the pandemic crisis. Further, we see some growth of the overall news output at the very end of the analysis period, which can plausibly be attributed to the, then, massive rise of Corona-related messages. Still, the stability of the overall volume for most of the analysis period is notable in light of the virtually unlimited publication space Facebook offers; so, the limiting factor is the production side (i.e., overall productive hours of the newsroom and probably also expectations of what the audience can and should plausibly digest, which in turn leads to specific rules and routines).

The relative share of Corona-related messages (as opposed to all the other, “regular” posts) also shows an interesting growth pattern, which also helps answer some of the critical questions regarding journalism’s performance during the early crisis. As mentioned before, some critics noted too much focus on the pandemic, with all the other topics being completely pushed back due to the virus crisis (Section \ref{introduction}). However, the data analysis does not fully support such criticism, at least for January and February. There was a first burst of news following the official announcement of the first infection in Germany late on January 27, and most media reporting this on the following day \citep{tagesschau.de_2020a, tagesschau.de_2020b}. However, there is no relevant impact on the overall news flow, and Corona-related news made up a minimal share of the messages, even more so in the subsequent weeks, as the situation remained contained to a few known and traceable cases in Bavaria. We see a second burst of news at the end of February when new chains of infection became known, which led to a phase of a difficult-to-trace spreading of the virus; accordingly, the message output reached a higher level. However, the Corona coverage was still not dominant at the beginning of March. With the first deaths in Germany on March 9 \citep{n-tv.de_2020}, we can observe a sudden increase of news output, and in the following days, the output of virus-related messages grew massively (for more detail, see Section \ref{topics_over_time}). 
 
\begin{figure}
  \centering
  \includegraphics[width=1.0\textwidth]{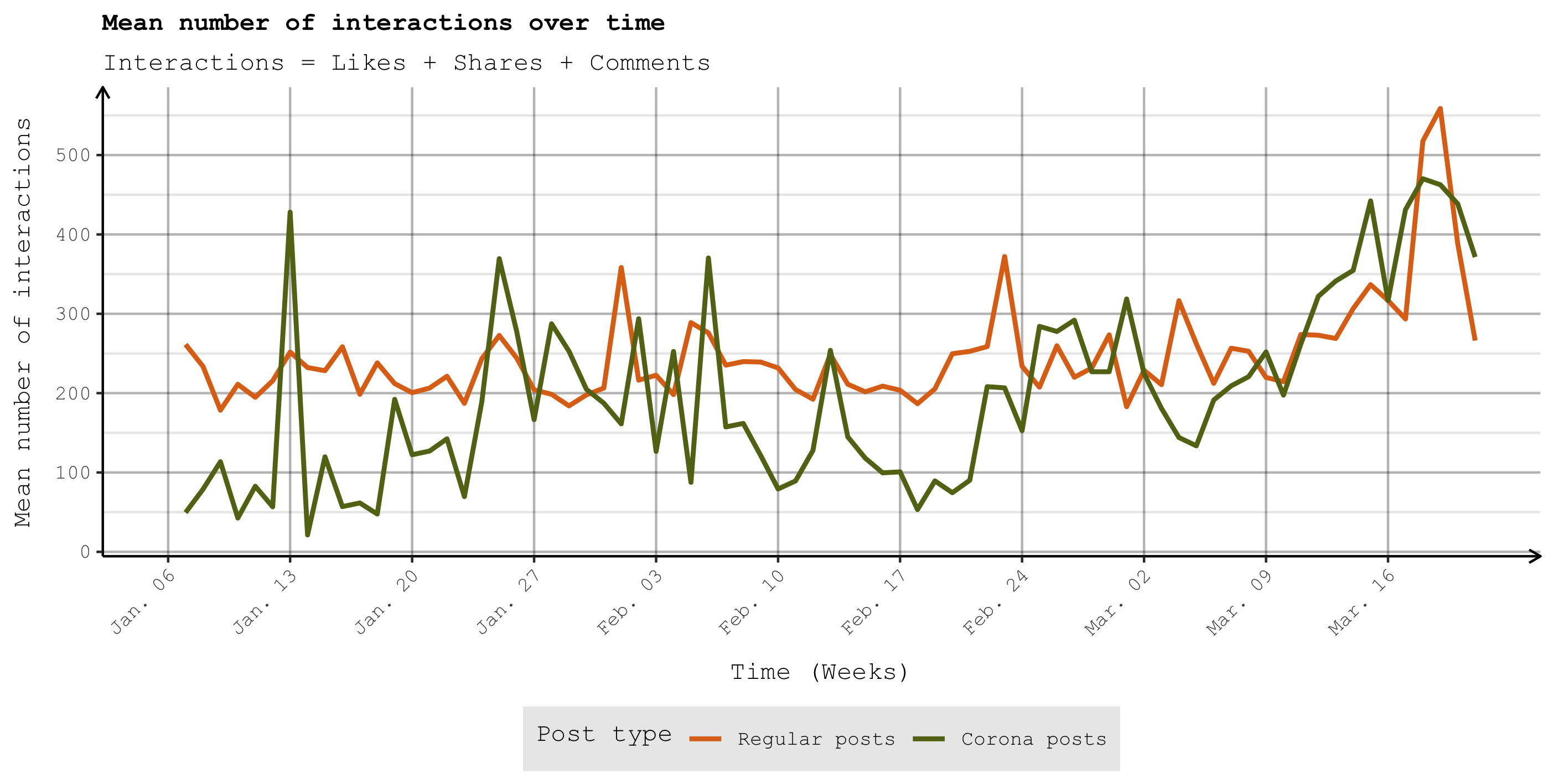}
  \caption{Mean number of interactions over time}
  \label{fig:interactions_over_time}
\end{figure}

\clearpage

Indeed, during that later phase of development, there is a “take over” of Corona-related messages, with a majority of messages having some connection to the virus. Here, we can also see the push-back of other news, as discussed by critics. However, there is some adaption of the system (i.e., in the form of the above-mentioned, slight increase of the overall news output), and still roughly one-third of messages were unrelated to Corona at the end of the analysis period.

Over the period of our study, all Corona-related posts generated a total of 5,732,155 interactions. It can be noted, however, that the number of interactions differs significantly among sites. Facebook pages from large national media outlets, such as Bild, Spiegel, or Welt, triggered an average of over 1000 interactions per post, while regional newspapers tended to receive between 100 and 200 likes, comments, and shares. Corona-related posts triggered an average of 391 interactions, compared to posts on other topics  with a similar average of 323 interactions. Looking at the interactions generated over time (Figure  \ref{fig:interactions_over_time}), a few days can be identified on which the Corona-related posts provoked a particularly large number of reactions. Especially at the beginning of January, the posts on Corona showed larger fluctuations due to the fact that there were hardly any posts on the topic at that time (see Figure  \ref{fig:posts_over_time}) and thus, individual posts from high-reach sites contributed to the spikes. For example, a post from Spiegel on the shutdown of Wuhan on January 25 reached almost 12,000 interactions. On the other hand, concrete events such as the death of the Chinese doctor Li Wenliang, who first discovered the new virus, on February 7, 2020, or the imminent closures of schools on March 15, 2020, triggered an above-average level of activity over several pages. All in all, however, media outlets did not—in contrast to what they were being accused of by some—trigger an excessive amount of user reactions and thus attention via the topic of Corona. 

\subsection{Identification of main topics}\label{main_topics}

One of the main interests in analyzing the performance of mainstream news during the Corona crisis is the topical composition of the news, answering the basic question of what journalistic news media were covering during the analysis phase (RQ2a). A topic modeling approach allows for an identification of the major topics found in the material. As mentioned above (Section \ref{methods}), we checked several solutions in terms of their mathematical and sense-making qualities. A 12-topic solution emerged as the most fitting structure for the material in both respects, allowing for an insightful interpretation (Table \ref{tab:topic_description}). 

\begin{table}
\small
 \caption{Topic description}
  \centering
  \resizebox{15cm}{!}{
  \begin{tabularx}{\linewidth}{
  >{\raggedright\arraybackslash\hsize=.04\hsize}X%
  >{\raggedright\arraybackslash\hsize=.15\hsize}X%
  >{\raggedright\arraybackslash\hsize=.2\hsize}X%
  >{\raggedright\arraybackslash\hsize=.49\hsize}X%
  >{\raggedright\arraybackslash\hsize=.06\hsize}X%
  >{\raggedright\arraybackslash\hsize=.06\hsize}X%
  }
    \toprule
    Topic & Description & Top terms & Example & Mean $\gamma$ & N, where $\gamma$ > 0.5 \\
    \midrule
    1 & Number of infected and dead & Germany, number, cases, Italy, rise, developments, China, infections, spread, infected & “Germany and France report many new infections with SARS-CoV-2, in Italy 49 more people die from the consequences of the virus. Meanwhile, Minister of Health Spahn addresses an appeal to the citizens.” (Spiegel) & 0.13 &  2052 \\ 
    2 & First chains of infection & case, infected, quarantine, county, tested, positive, confirmed, suspicion, patients, Heinsberg & “The coronavirus has most probably arrived in our region. Stephan Pusch, Administrator of the Heinsberg district, confirmed to our newspaper on Tuesday evening that two people were being treated for the virus in the Erkelenz hospital.” (Aachener Zeitung) &  0.12 &  1901 \\ 
    3 & Medical experts & virus, Germany, questions, experts, explain, answers, virologist, influenza, most important, pandemic, RKI & “Professor Christian Drosten is one of the leading virologists in Germany. In this interview, the professor from the Emsland region explains how many people in Germany might die from the Coronavirus and which measures would really be worth considering now.” (NOZ) & 0.10 &  1497 \\ 
    4 & Measures restricting public life &  measures, curfew, crisis, city, Bavaria, live, closed, public, spread, police & “In the fight against the spread of the coronavirus, the German government is asking the federal states to close a significant number of businesses.” (FAZ) & 0.09 &  1541 \\ 
    5 & International affairs \& consequences for travel & China, Trump, Wuhan, quarantine, USA, German, vaccine, Europe, passengers, flights & “The US denies foreigners access from China to stop the spread of the Coronavirus. Russia is even considering expulsion of infected persons. The government in Beijing is speaking of overreactions.” (Spiegel) &   0.07 & 1165 \\ 
    6 & School shutdown \& closure of borders &  school, closed, day care, parents, children, stay, must, students, Monday, home & “The schools are closed, teaching now takes place only digitally. However, there is strong criticism that the infrastructure in NRW schools is not sufficient for this. Our question of the day is: How does teaching and learning at home work out?” (WAZ) &   0.07 &  993 \\ 
    7 & Hoarding & supply shortages fear, disinfectants, hoarding, crisis, customers, patients, hospitals, medical doctors, police, pharmacies & “The hysteria surrounding the Coronavirus is spoiling the manners. First, supermarkets were hoarded empty. Now disinfectant material is disappearing en masse in hospitals.” (Ruhr Nachrichten) &   0.08 &  1196 \\ 
    8 & Economic consequences & crisis, companies, economy, consequences, Euro, Germany, pandemic, production, German government, Short-time work & ECB President Christine Lagarde reconsiders the caution of Europe's monetary watchdogs in the light of the Corona crisis: "There are no limits to our commitment to the Euro." (Berliner Zeitung) & 0.07 &  1155 \\ 
    9 & National cohesion & Merkel, crisis, comment, Germany, Spahn, Chancellor, fake, CDU, news, citizen & "This is an historic task, and it can only be accomplished together." (Donau Kurier) &  0.08 &  1236 \\ 
    10 & Culture  and sports (cancellations and new online events) & canceled, events, soccer, Bundesliga, Leipzig, big events, postponed, spread, take place, despite & “Federal Health Minister Jens Spahn recommends the cancellation of events with more than 1000 participants. Football matches and concerts would also be affected. Will participants be refunded for their tickets?” (KStA) & 0.08 &  1265 \\ 
    11 & Human interest (everyday life with Corona) & crisis, times, Osnabrück, help, region, information, for free, population, availability, offer & “[Name] is a high-risk patient himself. The 33-year-old suffers from Crohn's disease. However, this does not stop the young relaxation trainer from helping other people. Together with two other supporters, he has now founded the "[Location] Aid Initiative" during the Corona crisis. The aim is to link people seeking and offering help with each other.” (Focus) &  0.06 &  867 \\ 
    12 & Soft news & Italy, quarantine, foto, virus, Chemnitz, live, Saxony, times, city, video & “Bored at home? Not for Sarah Lombardi [German Pop Idol winner] and her new boyfriend! Those two know how to pass the time.” (Bild) & 0.05 &  720 \\ 
    \bottomrule
    \multicolumn{6}{l}{\emph{Notes:} LDA with \emph{k} = 12, \emph{$\alpha$} = 0.05, sampling method = Gibbs} \\
  \end{tabularx}\label{tab:topic_description}
  }
\end{table}

The spectrum of topics is wide, ranging from information about the pandemic and the spread of COVID-19 and the measures against the virus to the effects on various aspects of social life and human-interest stories. When compared to our previous analysis of alternative news media \citep{Boberg_Quandt_Schatto-Eckrodt_Frischlich_2020}, the topic structure is noticeably broader, giving a more complete overview of societal consequences, which is in line with expectations, as most alternative news media also consider themselves as heterodox and oppositional, therefore pronouncing problems and failures, and some of their long-term narratives, like migration, which did not play a central role in the journalistic “mainstream”.

The first two topics found in the messages under analysis are primarily information journalism, focusing on the latest data and the spread of the virus in a largely neutral manner. Topic 1 is present in somewhat more than 2000 messages and primarily contains reports on growing infection numbers and deaths. There is some focus on China and Italy here, as they were the sites of the two most relevant outbreaks in the analysis phase. This topic can also be linked to some of the criticized horse-race reporting (Section \ref{introduction}) where infection numbers of countries are compared like a competitive sports event. As our analysis shows, this type of coverage did indeed exist, so the criticism is not fully unfounded. However, the overall topic only accounts for slightly more than 10\% of the overall Corona-related corpus, and only a fraction of it is coverage in the style of horse-race reporting. So, for the full analysis period, the perception of this being dominant is not backed up by the data at all. It needs to be seen whether this may be still true for various phases of the coverage, though (Section \ref{main_topics}).

Topic 2 is similar to the first one in the sense of being fact-based news journalism; however, its focus is more domestic. Messages that contain this topic debate the first infection chains, predominantly within the country (with just a few messages discussing the early infection chains in China and, more generally, Asia). Many messages note the initial cases of Corona in the respective regions and where the infected individuals might have acquired the virus. It also includes stories on some of the initial centers of infection, like a carnival session in the Gangelt municipality in the district of Heinsberg, and how the virus was transferred via some super-spreaders in parts of the country.

The third topic gives voice to experts from the fields of medicine, virology, and epidemiology. This includes nationally present experts (Section \ref{topics_over_time}), but also a high number of local experts from local clinics explaining measures to contain the virus via hygiene and adapted behavior. Further, the topic also contains discussions of media appearances of experts, interpreting their statements, for example, on national talk shows. 

\clearpage

Topic 4 is present in a high number of messages. It is concerned with the implementation of anti-Corona measures in the cities, districts, and federal states, predominantly in a neutral style. Concrete information is provided on how the respective measures (closure of shops, playgrounds, etc., ban on visiting nursing homes) are actually implemented and how this is monitored by the police. It also contains questions regarding whether the people need to be prepared for even more drastic measures. 

The fifth topic focuses on the international situation again. It includes news on travel restrictions, often related to China and the USA, and reports of stranded passengers in hotel complexes or on cruise ships and repatriation operations. This “international affairs” topic also includes the widely discussed story that US President Donald Trump wanted to secure exclusive rights to potential vaccinations from the German company, CureVac, which also triggered political reactions from the German administration.

In contrast to this, Topics 6 and 7 primarily refer to domestic issues and societal reactions to the virus. The sixth topic contains information on school closure and the organization of emergency care. The main issues addressed here are the challenge of parents who now have to organize care, and past failures in the digitization of schools in Germany, which make them unprepared for a home schooling and online teaching. In addition, reference is made to alternative offers of digital teaching and education (e.g., public broadcaster’s offerings or the YouTube channels of third parties). This topic also encompasses the closures of borders, as both school and border closures are confounded in the analyzed material. Not only do they resort to similar words and concepts, they were also discussed during the same period.	
 
The seventh topic addresses reports on public panic reactions, including the hoarding of goods like toilet paper, food supplies, and masks. This includes discussions on how to prevent supply shortages, warnings of tricksters taking advantage of fear and panic, as well as concern about supply shortages in hospitals. While this topic revolves around the effects of COVID-19 related panic, and in that respect, is somewhat comparable to topics that were present in alternative news media during that period \citep[see][]{Boberg_Quandt_Schatto-Eckrodt_Frischlich_2020}, it also contains ample warnings against disproportionate reactions and anti-social behavior, and therefore, it is quite different in message content and tone (i.e., it is reporting issues of panic, but does not intentionally aim at panic mongering).   
 
The eighth topic is comprised of a discussion of the economic damage the Corona measures have. Industry representatives also have their say here and report on their problems. Comparisons are made with the financial crisis and its economic effects. Overall, this topic primarily focuses on damage from a system perspective, and most messages are phrased in a neutral style, with a few exceptions that use more direct emotional appeals (e.g., in Spiegel Online: “The coronavirus threatens human life and the hyperlinked economy. What are the consequences for jobs, trade flows and our prosperity?”). 

The ninth topic is already known from the previous analysis of alternative media \citep{Boberg_Quandt_Schatto-Eckrodt_Frischlich_2020}; it’s about “national cohesion” in a time of crisis. Many posts that include this topic refer to the public speech of Chancellor Angela Merkel on Corona and the counter-measures to limit its spread. The majority of messages reacts to it in a positive way, highlighting Merkel’s empathy for the people in difficult times, with just a few critical exceptions. This topic also includes messages commenting on fake news and conspiracy theories, and the dangers of these for national cohesion and the success of anti-Corona measures. Partially, this topic can be seen as proof for “political parallelism” \citep{Hallin_Mancini_2004} in the coverage, which argues close to political actors, and calls for a support of the administration’s goals in order to retain national cohesion in the crisis \citep[which is a variant of the “rally around the flag” effect; see][p. 11]{Boberg_Quandt_Schatto-Eckrodt_Frischlich_2020}. Some critics also labeled this “system journalism” \citep{Jarren_2020}, in line with a recurring narrative in alternative news media regarding traditional “legacy” media being part of the power elite and “the system.” However, the topic only accounts for roughly 7\% of the Corona-related messages during the analysis period, so this is hardly dominating the message output. Again, a further inspection of the time-based development is necessary to see whether this holds true for the various phases of the topical flow.

Topic 10 contains information regarding cancelled and postponed events, especially in the cultural sector. Some of this is service journalism also offering tips on what to do when staying at home during the crisis or what type of online events and streams (e.g., concerts) might be worth watching. Further, this also includes coverage on sports events, including “ghost games” (with no spectators in the stadium) and how the Corona crisis affects the Bundesliga (national soccer league) and its (then ongoing) season.

Topic 11 focuses on everyday life with Corona and describes this along with individuals, often in their local settings. Messages in this category portray people who either suffer particularly from the pandemic (economically or in terms of health) or people who start initiatives against its social effects  (e.g., aid organizations or school children who shop for the elderly, etc.). In addition, this “human interest” category also includes forms of service journalism that answers everyday “practical” questions such as, "Where do I put the green waste when the collection points are closed?”

Finally, Topic 12 is the least coherent one, uniting a broad range of soft news and boulevard topics. On the one hand, there are reports about celebrities, animals, or Internet finds, such as "handshakes in times of Corona," often accompanied by clickbait headlines like "this video goes viral now." On the other hand, there are highly emotional posts, often depicting individual fates and reporting on sick or dying relatives or divided families. Similar forms of attention-grabbing clickbait material were also present in the messages from alternative news media during the same analysis period. However, the respective topic’s share of the analysis corpus was higher than what we find here; about 6\% of alternative media messages could be attributed to the respective topic versus 4\% in the current analysis; furthermore, another topic where Corona was just used as a buzzword could be found in more than 3\% of the alternative news messages, something we couldn’t find as a coherent topic in the current analysis.

Summing up, the topics present in the Corona coverage of the journalistic media on Facebook cover a broad range of pandemic related messages, from pure information and news journalism, medical background information and expert voices, criticism regarding economic effects and school closures, to service journalism, local information and human-interest stories. Some topics include elements of the heavily criticized forms of horse-race reporting and uncritical “system journalism” or “court circular” (Section \ref{introduction}), but their share of the overall corpus is limited. 

However, it is interesting to see what we don’t find here, especially when contrasting the findings with our previous analysis of alternative news media. The topics, as inductively generated by the LDA, did not contain the various forms of outspoken anti-systemic messages as identified for the alternative media \citep{Boberg_Quandt_Schatto-Eckrodt_Frischlich_2020}, like the many portrayals of “worst case scenarios,” “fear and panic inducing messages,” criticism of the “failures of the government to prepare,” and comments on why the “measures are not reasonable” or the “chaotic crisis management.” Indeed, a direct comparison shows that the alternative news media primarily consists of fundamental opposition to political or official actions. Their general narrative is, broadly speaking, anti-systemic, that is, not only disagreeing with the administration, but also a perceived “overall system of oppression” that extends to the very basis of the existing liberal democracy. In contrast, journalistic news media try to represent multiple perspectives on the various parts and constituents of society (and therefore operate within the existing social and political system). In that sense, some of the criticism of journalism’s performance during the (early) crisis can be explained by the observers’ own expectations; some may have hoped for a more fundamental “external” view, as an opposition to the administration and authorities, which is not what happened on a broad basis during the analysis period. Again, criticism regarding the effects of Corona-related measures is indeed present in the material, and even in a notable number of messages, but it is not based on a fundamental disagreement with the political and social reactions (as it is in the case of alternative news media). However, a closer look at the emergence and development of topics during the early pandemic may also help in understanding why some of the criticism may be true for some phases and inadequate for others. 

\subsection{Emergence and development of topics over time}\label{topics_over_time}

\begin{landscape}
\begin{figure}
  \centering
  \includegraphics[width=21.5cm]{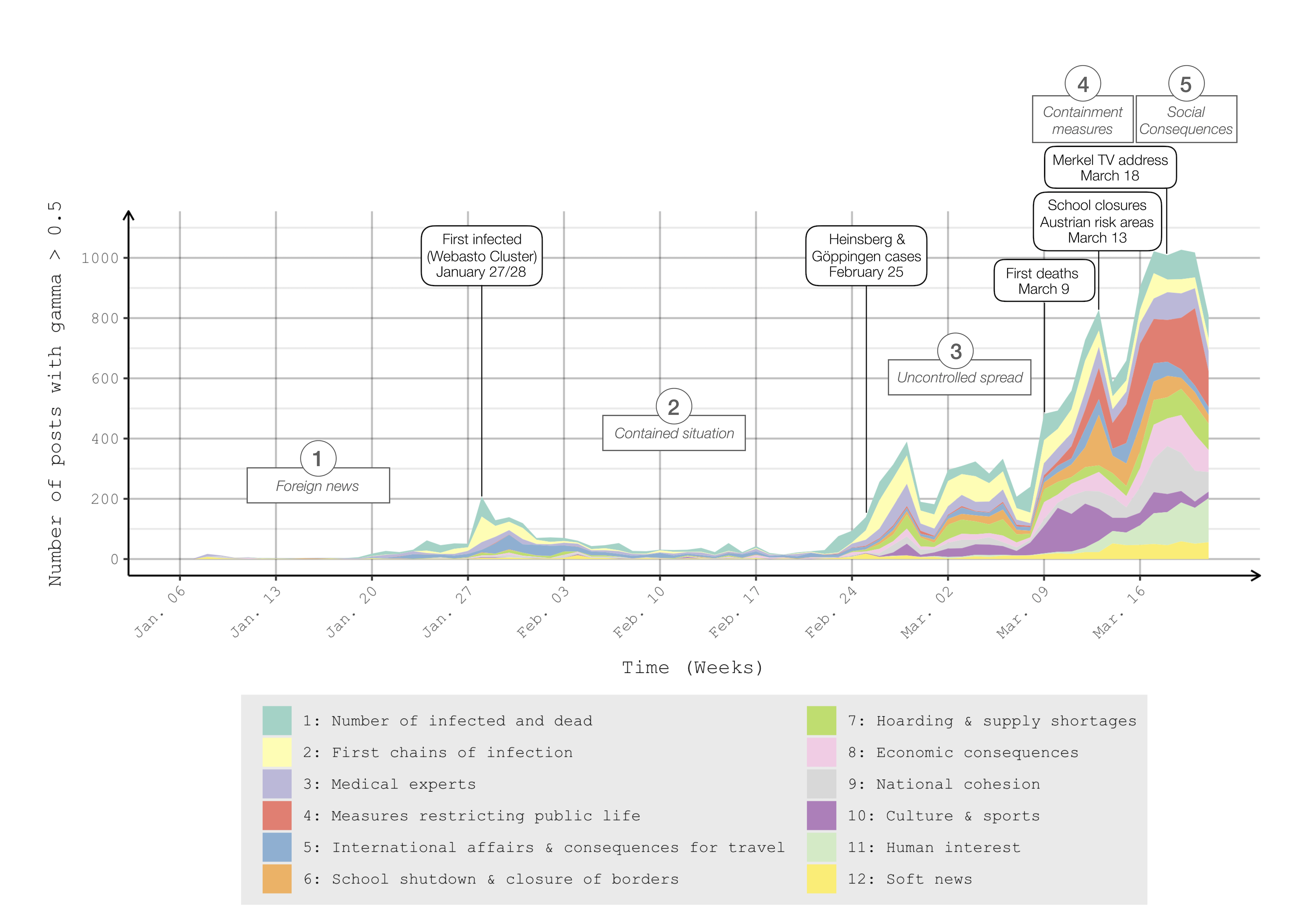}
  \caption{Number of topical posts over time}
  \label{fig:topic_over_time}
\end{figure}
\end{landscape}

The previous discussion of topics was based on the overall corpus so far, and ignores the publishing dates of the messages. As mentioned there, some questions, especially regarding criticism of news media’s performance in the crisis, depend on phases in the news flow, and this also refers to our RQ 2b. The previous rough overview of the news flow in the early crisis revealed bursts and spikes in the output, and a more fine-grained analysis is helpful in identifying phases with varying message composition (Figure  \ref{fig:topic_over_time}). 

First, it is notable that the overall shape of the curve reminds the respective earlier analysis of the alternative news media’s output \citep{Boberg_Quandt_Schatto-Eckrodt_Frischlich_2020}, with somewhat more pronounced weekend drops (the difference is especially obvious on March 7 and 8, where the alternative media didn’t have such a strong production dent). The overall fit of the shapes of the mainstream journalistic news media and the alternative news media is interesting in so far as both media experts and alternative media criticized the performance of journalism for either a (first) sluggish or (then) exaggerated reaction. However, the same criticism would equally apply to alternative media themselves, as their reaction time and relative volume changes do parallel what we see in the current analysis. Indeed, the reactions of both types of media are very much following external events (as will also be discussed in detail in the subsequent analysis). Therefore, it is plausible that they were only partially steered by the actors or news organizations themselves, if at all, so that news bursts \citep{Buhl_Guenther_Quandt_2018, Buhl_Guenther_Quandt_2019} are, grosso modo, not “controlled” by the media (in the sense of conscious and individual power over the process), but they are rather emergent phenomena, based on systemic logics. 

The close ties between specific events and developments in society on the one side and the journalistic Facebook output on the other becomes even more evident when looking at the topic composition over time. Here, we can also observe stark differences to our previous analysis of alternative news media \citep{Boberg_Quandt_Schatto-Eckrodt_Frischlich_2020}, that is, while the overall response time and volume changes are comparable, the actual content of the responses is mostly not, with some small exceptions.  

\emph{Stage 1: Foreign news \& not our problem}.\quad Indeed, the first stage of the virus reporting is one of these exceptions, as it’s exactly the same as in the alternative news media—it is simply a non-response, as we wrote elsewhere: “Up to January 20, the virus was essentially a non-issue in the analyzed Facebook debates.” \citep{Boberg_Quandt_Schatto-Eckrodt_Frischlich_2020} Beyond a few singular mentions, the first outbreak of the virus in Wuhan, China remained unnoticed in German media. This changed slightly with the initial infections outside China in Japan, Thailand, and South Korea, and the first WHO situation report, dated January 21 \citep[for an example of an international timeline see][please note that there may be +/-1 day differences in various circulating timelines, due to time-zone differences]{CNN_Editorial_Research_2020}. Some media reported infection numbers and the first infection chains were reported, but the overall output remained limited, as this was largely considered a “foreign news” item. 

\emph{Stage 2: First domestic cluster -- contained}.\quad Things changed suddenly on January 28 when the first known infection in Germany became widely publicized \citep{tagesschau.de_2020b, tagesschau.de_2020a}: The arrival of the virus “at home” made this a domestic news topic, and the response was a first news burst on that day. Unsurprisingly, the “infection chains tracing” topic is very pronounced here. This, and the spike in Topic 1, can be interpreted as an initial journalistic reaction with “breaking news” style messages focusing on facts and numbers (which are present in these two topics). In the days following the first spike, we can observe some expert voices in the output (Topic 3), as well as a discussion of international affairs (Topic 5). The latter is also partially explained by the first cases being employees of automotive supplier, Webasto, in Bavaria, with an infection chain that could be traced back to a visiting colleague from China (which made this both a domestic and foreign story simultaneously). This “follow-up” phase after the first spike contains more interpretation and explanation of events, which is also plausibly explained by the logics of journalistic production (i.e., initial information and news coverage is real-time production based on events reporting and immediately available facts, while interpretation and explanation takes more time, as it requires background research, reflection, and identification of ambiguous information). As the situation seemed to be traceable and contained \citep{Robert_Koch-Institut_2020a}, the interest in the topic slowly diminished. In the following days, we only saw limited reporting, with just a few smaller bumps in the output \citep[e.g., following the first death in Europe, a Chinese tourist in France, on February 15;][]{tagesschau.de_2020c}. 

\emph{Stage 3: The uncontrolled spread and supply issues}.\quad Things changed considerably at the end of February; infections in the area of Heinsberg and the city of Göppingen became public on February 25 \citep{Sohr_dpa_2020} and marked the start of a wave of infections, where tracing became more and more difficult. The first infected individual in Göpping seemingly had acquired the virus during a trip to Italy, where the situation was already much more difficult. In particular, the Lombardy region was heavily affected by an uncontrolled virus spread, which led to Germany’s decision to declare Italian regions as risk areas on February 26 and 27 \citep{Robert_Koch-Institut_2020b, Robert_Koch-Institut_2020c}. These events are paralleled by a sudden growth in the coverage, with stories on the infection chains and reporting of infection numbers as the imminent reaction, but also a high number of messages that contained expert voices, in particular virologists, epidemiologists, and other medical personnel. Some other topics also started to emerge during this period; the topic of hoarding and supply shortages (7), debates on the economic (8) and cultural (10) consequences, and the first signs of the “national cohesion” (9) however, not in a very pronounced way. It is interesting to see what’s also missing from the topical composition during that period; international affairs and consequences were mostly not present, which is indicative of the domestic nature and framing of this outbreak as the “virus arriving at home,” and a clearly “inward” perspective, despite COVID-19 affecting many other European countries during that time and emerging worldwide effects. After a weekend dent (February 29 and March 1), one can observe a similar topical composition, with some relatively limited international and travel related messages (i.e., Topic 5) added to the mix. 

\emph{Stage 4: First deaths and growing containment measures}.\quad Directly after the following weekend dent (March 7‒8), there was a rather explosive growth of output, which is in sync with the events that unfolded on March 9 and subsequent days. On Monday, March 9, two people died from Corona in Germany \citep{n-tv.de_2020}, and the unchecked spread of the virus led to debates on stronger restrictions of public life to contain the threat. This was paralleled by the emergence of a corresponding topic on measures restricting public life (4) and the growth of debates on shutting down schools and closing borders (6), as well as a slight rise in reporting on potential economic consequences (8). However, the most obvious growth can be seen in messages occupied with the impact on culture and sports -- in particular, ghost games, and the potential halt of the German soccer league (Bundesliga), which was officially decided on March 13 \citep{dpa_2020}. The acceleration of events that unfolded in the week of March 9‒15 culminated in a remarkable burst of messages on school shutdowns and closures of borders, as condensed in Topic 6. In particular, numerous German ski tourists returning from Austria showed signs of COVID-19, followed by the declaration of Austrian state of Tyrol as risk area by the German Robert Koch Institute \citep{Robert_Koch-Institut_2020d} on March 13 and the full quarantine of some ski areas by the Austrian government on the same day \citep{Dahlkamp_Goos_Hoefner_Hutt_Latsch_Lehmann_Mayr_Polonyi_Stock_2020}. In parallel, a majority of German states also decided to close schools; the Saarland had already decided on the in the night of March 12, with most states following on March 13, effective from the beginning of the next week \citep{tagesschau.de_2020d}. Unsurprisingly, this was instantaneously picked up in the coverage, as the effects on the people were immediate and direct. 

\emph{Stage 5: Social consequences and rallying calls}.\quad After another weekend dent, we saw a further change in the news composition. The discussion of the now tangible measures (4) received much more room, and, to a lesser degree, also expert voices (3), while other topics somewhat faded from the spotlight (like the sports and culture related messages). Also, economic consequences got more coverage (8), and there was a notable growth in the “national cohesion” topic (9). There are multiple explanations for this; part of this may be due to a “rally around the flag” effect, part of it can be attributed to the national television address of Chancellor Merkel on March 18 \citep{Tagesschau_2020}. Interestingly enough, this very late emergence of such a topic somewhat counters the criticism that mainstream media were too system-affirmative, at least for the overall time period; and indeed, the corresponding topic was actually more pronounced in the alternative news media \citep[who themselves chastised the journalistic media for being too close to “the system”;][]{Boberg_Quandt_Schatto-Eckrodt_Frischlich_2020}. Finally, during the last phase of the news flow, two other topics became relevant—human interest stories (11) and soft news (12). This is a reflection of the events now being close to the everyday experience of the average person, which was picked up by the media.

\begin{figure}[h!]
  \centering
  \includegraphics[width=0.6\textwidth]{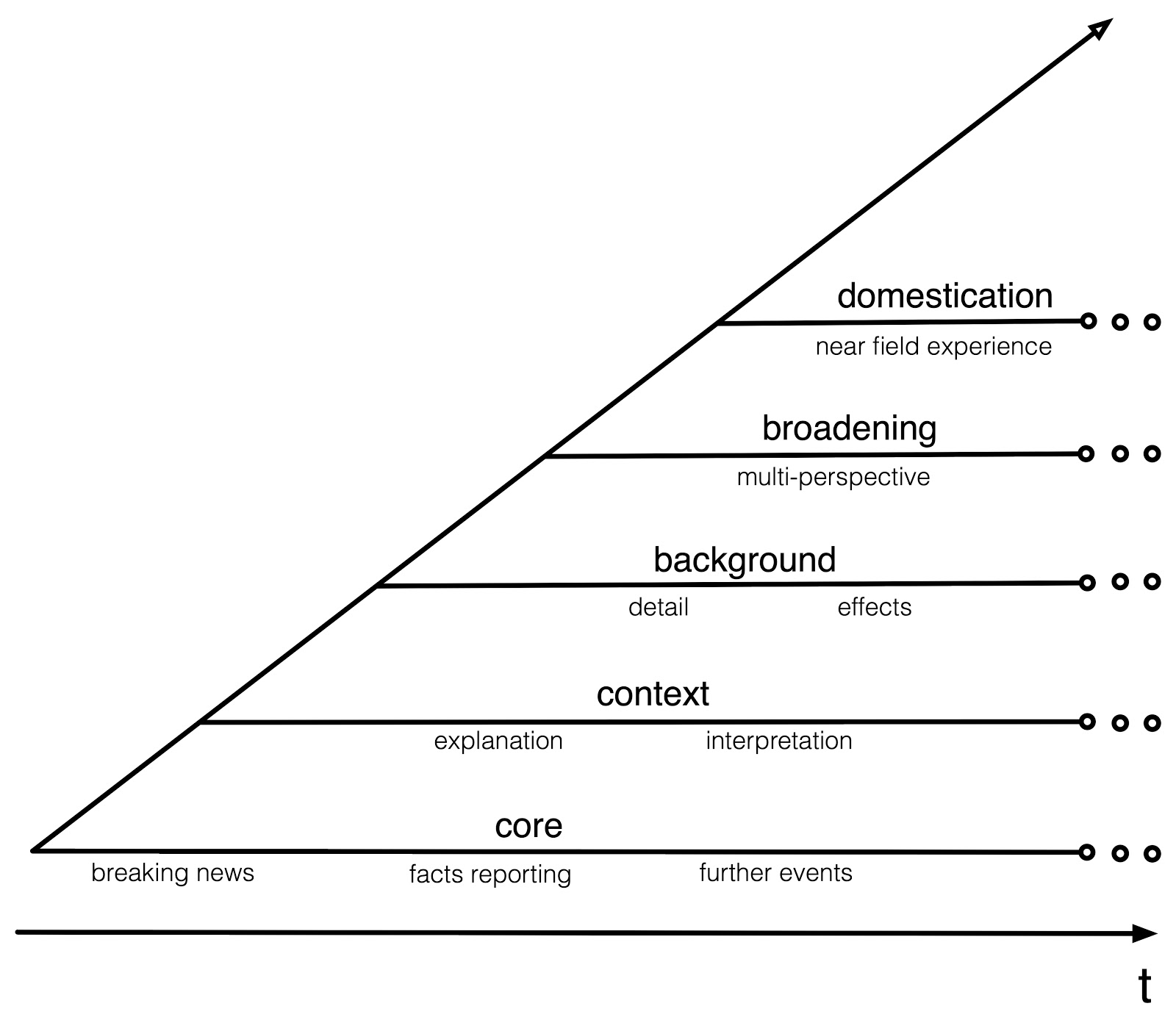}
  \caption{Generalized development of news flow in response to critical events}
  \label{fig:generalized_development_of_news_flow}
\end{figure}

Overall, this time-based analysis of the topical flow reveals changing patterns in the news composition during the various stages of development. News bursts and sudden changes can be explained by events and journalistic reactions to them, while pronounced dents are primarily the mundane effects of weekend production drops or stops. What can be observed here are emergent phenomena, which are not based on direct control of the news flow by the media, but rather of journalistic “steering” in a maelstrom of pandemic related events (Section \ref{new_platforms}). On the other hand, this steering means that the patterns are not random. The initial reactions to the “frame breaking” events seem to contain primarily news and information journalism (i.e., breaking news material and immediate facts reporting), with follow-up explanatory and interpretative phases to add context, then additional background information, and, in the long run, a broadening of the topical structure to offer a multi-perspective view on the varying societal fields, from sports, culture, and education to the economy. Further, domestic and close-to-home stories of the effects on the average citizen appeared in the last stage of the development, widening the coverage to the near-field experience of the users. The sequence of (re)actions of journalism follows a plausible general logic (displayed in Figure  \ref{fig:generalized_development_of_news_flow}), both on the basis how events unfold over time and how news organizations need to organize their workforce and routines. It needs to be noted that these generalized reaction patterns are not necessarily identical with the stages of the pandemic coverage as described above as some of the phases can also happen during one and the same stage, and as the process can also stop without reaching the full spectrum of responses. In particular, if the event doesn’t have enough impact on the overall society, the later phases won’t be very pronounced (e.g., as it was the case with regard to the first German cluster, which spiked just briefly, but then the coverage died down).

Future research will need to test whether such a sequence of coverage phases could be observed in other crises beyond the pandemic; however, it seems to be plausible that there is some internal logic to this, and that the order of these phases also depends on journalistic rules and routines, as discussed in previous literature (Section \ref{crisis_coverage}). 

\subsection{Actors}\label{actors}

\subsubsection{Top-ranked actors}\label{top_ranked_actors}

As discussed in our previous working paper \citep{Boberg_Quandt_Schatto-Eckrodt_Frischlich_2020}, news media, especially in the political realm, use personalization as a tool for reducing complexity and telling stories via individual actors \citep{Bennett_2012, Boumans_Boomgaarden_Vliegenthart_2013}. Beyond individual persons, actors can also be groups, institutions, and nations, which are portrayed as a unified entity, and much like persons, are supposed to act in a deliberate and coordinated way. Such personalization is part of the normal functioning of journalism, as it helps in making a complex world more tangible for its audience. However, critics noted a very limited focus on very few key figures in the crisis, and a heroization of seemingly omni-present individuals, for example, leading medical experts and virologists (Section \ref{introduction}). 

For the current analysis, following our research interest (RQ 3a) and allowing for a comparison with our previous analysis of alternative news media from the same overall corpus \citep{Boberg_Quandt_Schatto-Eckrodt_Frischlich_2020}, we identified the most frequently named entities \citep{Manning_Surdeanu_Bauer_Finkel_Bethard_McClosky_2014}, including organizations and institutions (i.e., the government, parties, the WHO, etc.). To streamline the analysis, we excluded individual nations as actors, as these were not central to our study (also in line with our previous research). 

\begin{figure}
  \centering
  \includegraphics[width=1.0\textwidth]{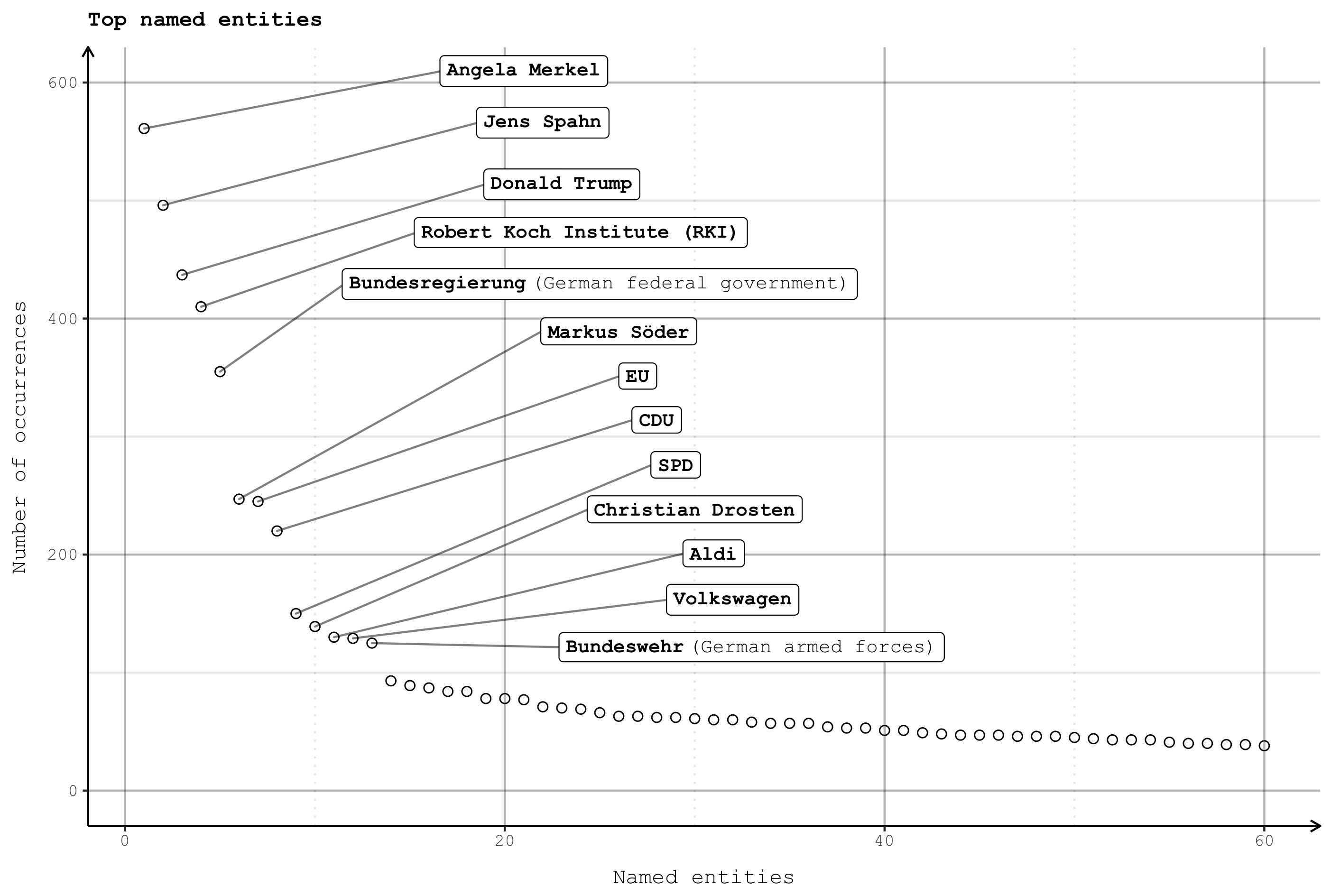}
  \caption{Number of occurrences of actors, figure restricted to top 60 of 1,895 names entities}
  \label{fig:ner}
\end{figure}

The analysis reveals 1,895 identifiable, named entities in the material \citep[as opposed to 421 named entities for alternative news media, using the same identification method; see][]{Boberg_Quandt_Schatto-Eckrodt_Frischlich_2020}. While this speaks for a large variety at a first glance, further inspection of the mentions per entity reveals a power law distribution with stark differences between the frequently named top entities and a long tail of actors only showing up a few times, or even just once (Figure \ref{fig:ner}; this includes the first 60 named entities, leaving out the long tail for easier readability). However, despite the strong concentration, this is not sufficient proof that the criticism of heroization and limited court circular in support of the government (Section \ref{introduction}) is appropriate. Indeed, such a focus on central actors has been found in previous analyses as well, even outside crisis coverage \citep[e.g., for the German online coverage of the elections, see][]{guenther_domahidi_mediale_sichtbarkeit_2017}. Therefore, it is also crucial to have a look at the identity of the top actors.   

The top actors are Chancellor Angela Merkel (561 mentions), followed by the German Minister of Health, Jens Spahn (496), US President Donald Trump (437), the Robert Koch Institute, as the central federal government agency and research institute announcing risk areas, infection numbers, and the effects of the virus (410), and the German government (355). These actors' appearance on the top list is no surprise, as they were, indeed, very present as central actors and decision makers in the crisis, and they also ranked high on the list of the alternative news media \citep{Boberg_Quandt_Schatto-Eckrodt_Frischlich_2020}. Actually, it is quite remarkable that while Merkel leads the list, she does not stand out as much as she did in the coverage of alternative media. There, she had more than two times as many mentions as every other actor, and the difference of the top political actors as compared to other entities was more pronounced. Some of the political elite persons, for example, Peter Altmaier  and Olaf Scholz, showed up nearly as frequently in alternative media as they did in mainstream media despite the overall number of posts in alternative media being more than seven times lower (Altmaier had 25 mentions in alternative and 40 in journalistic mainstream media; Scholz had 20 in alternative and 34 in journalistic mainstream media). In that sense, the mainstream media were much less preoccupied with these central political actors and the government than the alternative media—so, the criticism of court circular fits the alternatives more, albeit in a twisted negative form (which is somewhat ironic since they typically criticize the mainstream media for publishing court circular as well). 

Other groups of prominent actors include more top-ranking politicians (Markus Söder, 247 mentions; Horst Seehofer, 93; Armin Laschet, 78; Friedrich Merz, 78), the two government parties (CDU, 220; SPD, 150), international institutions (EU, 245; WHO, 89). We also find the Bundeswehr here (i.e., the army, which was used as support in heavily affected areas; 125). Further, the top list also includes German virologist Christan Drosten (with 139 mentions), who served as an advisor to the government in the crisis. The prominent role of Drosten in public was, indeed, criticized as heroization of virologists in the pandemic. However, he is the only such person showing up in the top list of most frequently mentioned actors, and it needs to be mentioned that he could be found on the top list of alternative news media as well \citep{Boberg_Quandt_Schatto-Eckrodt_Frischlich_2020}. This also reflects his frequent appearance in public, for example, in press briefings alongside the Minister of Health, Spahn.

Finally, companies form a relevant group of named entities in the top list include Aldi (130 mentions), Volkswagen (129), CureVac (84), and Webasto (84). All of these were tied to specific stories or represent a certain aspect of virus-related effects, so their appearance in the list seems logical. It is interesting to see, though, that none of these appear in the top list of alternative news media in our previous analysis. Instead, we could find many more individuals as actors there, including multiple international leaders, like Erdogan or Macron, as well as most oppositional parties \citep[with the German right-wing party AfD ranking sixth, which massively overdraws its role during the analysis phase; see][]{Boberg_Quandt_Schatto-Eckrodt_Frischlich_2020}. While these persons and parties also appeared in the list of named entities of mainstream journalistic media, they didn’t reach the twenty most prominent actors. So, when comparing the two types of media, it becomes apparent that the personalization is actually higher for alternative news media, as is the preoccupation with political processes and power issues, while the traditional journalistic media paint a more diverse picture, including other societal actors, like companies and institutions. 

\subsubsection{Co-occurrence analysis}\label{co_occurrence_analysis}

\begin{figure}[h]
  \centering
  \includegraphics[width=0.9\textwidth]{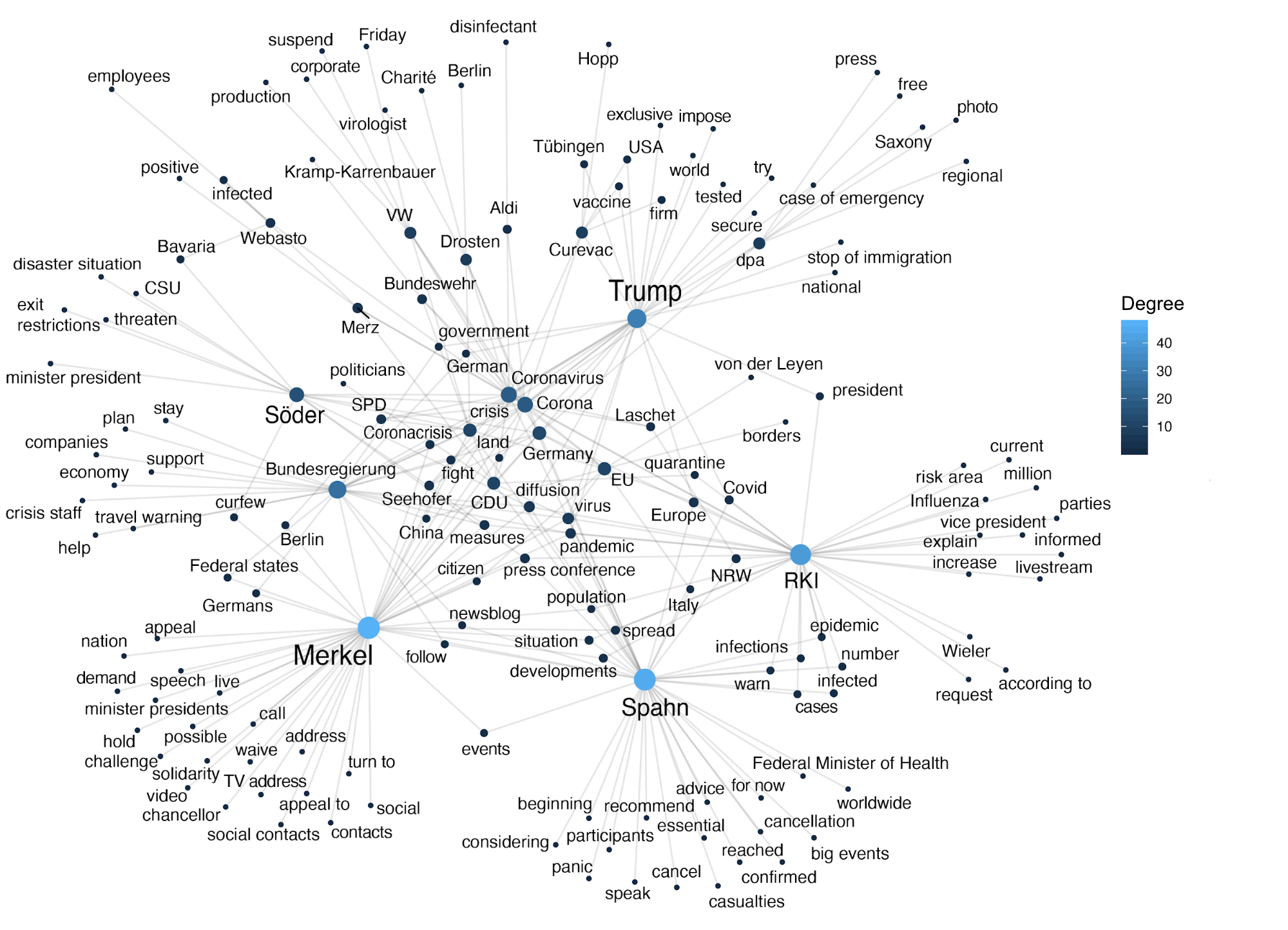}
  \caption{Co-occurrences of most frequent named entities, \emph{N} = 21019, filtered by edge count > 15}
  \label{fig:occ_ner}
\end{figure}

The analysis of named entities already reveals some of the focus in the Corona coverage of news media on Facebook via the top actors we found in their messages. A co-occurrence analysis helps in identifying in which context the identified actors appear (RQ3b). The main actors are positioned in relation to the terms that are most frequently connected with them through co-occurrence in the same messages, forming a graph of terms\footnote{Some entities seem to appear twice in this graph. This is due to the fact, that terms were translated from German post-analysis, that is, after the calculation of the graph. In some cases two (or more) distinct German terms only have one corresponding English term.} (Figure  \ref{fig:occ_ner}). The network gives insight into which actors have a high degree, how central they are in relation to other terms, and whether they are placed in densely connected sub-networks, which allows for an analysis of the meaning construction through relations of terms.

Through visual inspection of the co-occurrence graph, we can directly identify the most central and connected actors, which are also the most frequently named entities: Chancellor Merkel, Minister of Health Spahn, American President Trump, the Robert Koch Institute (RKI), the government (Bundesregierung), and the Bavarian Minister President and leader of the co-governing Christian Social Democrats (CSU), Söder. Most of them are central to relatively well-defined sub-networks, with only limited overlap to other sub-structures. This partially reflects a division of tasks and the roles the German governmental and official actors played during the crisis. Chancellor Merkel is primarily associated with calls for solidarity and her public address to the nation, that is, a representative and moral function. Spahn and the RKI were frequently informing the public in press conferences, announcing infection numbers, discussing health issues and the measures against the virus, as well as giving recommendations (with some sub-division of Spahn being more closely connected to recommendations and advice, while the RKI focused more on the medical aspects and the international spread).

Söder, on the other hand, is more closely connected to restrictions and measures taken against the threat (i.e., representing a risk and safety perspective), while the government is linked to crisis management, travel warnings, and curfews (i.e., rather operative issues). Another central actor, US President Trump, is prominently linked to a specific story about him trying to exclusively secure the rights for a potential vaccine from German company, CureVac, for the United States, which caused an intervention of the German government. Further, he appears in the context of an immigration stop to limit the virus spread. 

Like CureVac, most other companies that appear in the data set are either tied to a specific story or represent a particular aspect of the crisis. Employees of Bavarian automotive supplier, Webasto, were the first infected in Germany, while supermarket chain, Aldi, is mentioned in the context of hoarding (and a sale of disinfectants, which caused some chaotic scenes). Volkswagen represents the economic effects, and is tied to the halt of car production (and more generally speaking, the disruption of the normal functioning of the industry).

Summing up, the co-occurrence analysis allows for more detail in understanding how the various actors were portrayed in the crisis. On the one hand, we can observe a contextualization with regard to specific stories and actions of the analyzed entities. This could be regarded as the event-based, reporting perspective on the news (i.e., asking “what happened?”). On the other hand, we can also see how the actors were linked to more general topics or themes, and how they were utilized as representatives for various aspects of the crisis or even whole political or societal fields. By examining the data from this perspective, we learn about the framing of these actors in the news and how they became components 
in the construction of news reality (i.e., asking “What do these elements represent?”). 

When comparing the current findings with the previous analysis of alternative news media, we can observe a more holistic view on the functioning of a complex societal system and a multi-perspective on a variety of actors and contexts, which is a stark contrast to alternative media’s relatively unitary anti-establishment, anti-system narrative. Still, the mainstream media also tried to reduce the complex, and sometimes contradictory, reality by depicting actors in specific roles and reducing them to “typifiers” of one story, a specific aspect of the events, or a political or social field. While this can be criticized as reductionist, which was the case during the crisis (Section \ref{introduction}), it may also be understood as a necessity of the journalistic process to reduce the complex events to a workable and coherent tale of the world. 
 
\subsection{Sentiments}\label{sentiments}

As noted above, there has been criticism of the mainstream media being predominantly affirmative of the political decisions and systemic reactions during the crisis—and such criticism was also coming from the alternative news media, as some of them perceive the journalistic mainstream as part of “the system” \citep[see][]{Boberg_Quandt_Schatto-Eckrodt_Frischlich_2020}. In our preliminary analysis, we will focus on one aspect of this: the sentiments (RQ4) in the respective posts, which can be regarded as an indicator of the evaluations as expressed in these messages. In particular, we will analyze the level of negativity in this working paper, using a sentiment dictionary for German language (Section \ref{analytical_approach}).

A first analysis of the average prevalence of negative sentiments (Table \ref{tab:sentiment}) shows a relatively small difference between the scores for Corona-related posts and the overall prevalence; the messages that deal with the virus and its spread are more frequently negative. This is not surprising, given the fact that the language relating to a pandemic already contains numerous negative terms like “crisis,” “infection,” or “epidemic.” Other terms contributing to this also include “sorrow,” “fear,” “fight,” “threat,” “danger,” “hard,” or “sick,” to name but a few. So, in such a language related analysis, a pandemic is truly negative by definition. 

However, the comparison with alternative news media, using the same type of analysis, shows some interesting differences. For the alternative media, the difference between the score for Corona-related posts and the “baseline” score for all the messages in the respective corpus is clearly larger; their Corona messages are predominantly negative. Further, we can see that the coverage in the alternative media contains more negative messages per se, so the crisis amplifies an already quite negative language. The differences between the two media types are extremely significant, which does not surprise, given the numbers are based on a substantial sample. 

\begin{table}[h]
 \caption{Prevalence of negative sentiments in posts of mainstream and alternative news media}
  \centering
  \begin{tabularx}{\linewidth}{
  >{\raggedright\arraybackslash\hsize=.22\hsize}X%
  >{\raggedleft\arraybackslash\hsize=.29\hsize}X%
  >{\raggedleft\arraybackslash\hsize=.29\hsize}X%
  >{\raggedleft\arraybackslash\hsize=.1\hsize}X%
  >{\raggedleft\arraybackslash\hsize=.1\hsize}X%
  }
    \toprule
    Negative sentiments in & Mainstream news media & Alternative news media & t-value & p-value \\
    \midrule
    Regular posts & 0.38 & 0.48 & 19.359 & 0.000 \\
    Corona-related posts & 0.42 & 0.57 & 13.302 & 0.000 \\
    \bottomrule
 \end{tabularx}
 \label{tab:sentiment}
\end{table}

\begin{figure}[h]
  \centering
  \includegraphics[width=1.0\textwidth]{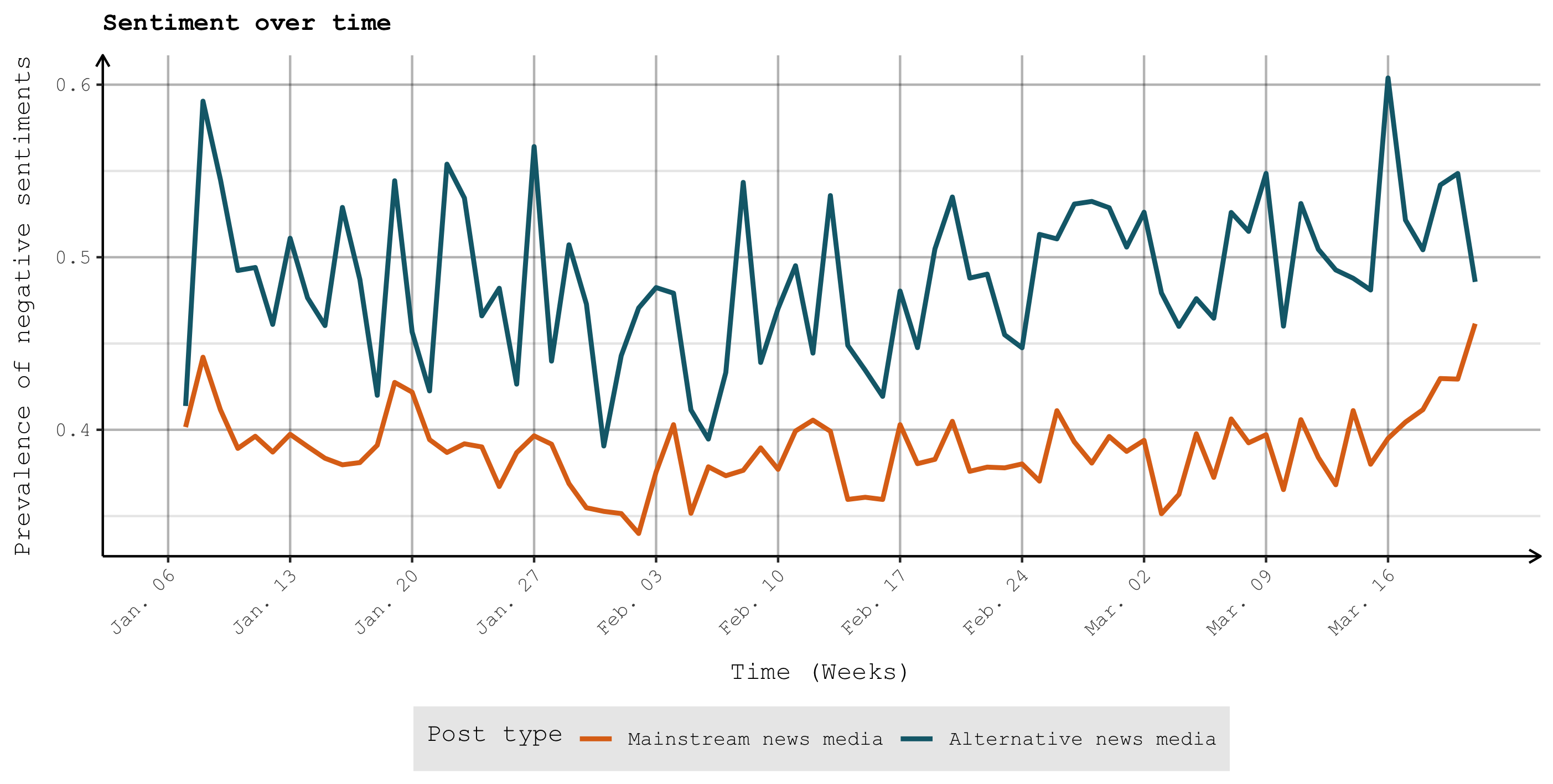}
  \caption{Negative sentiments of Corona-related posts over time (\emph{N} = 20,497)}
  \label{fig:sentiment_over_time}
\end{figure}

An additional time-based analysis (Figure  \ref{fig:sentiment_over_time}) reveals these differences to be present throughout the entire analysis period. In addition, we see notable fluctuation, especially for alternative news media, but no massive overall trend. The values for February are relatively low, with high values at the beginning and the end of the observed period. Naturally, this first data exploration has to be taken with a grain of salt, as it only allows for a very broad estimation of negativity in the coverage. However, in this “broad stroke” analysis, the differences between the two media types are quite pronounced and cannot be just explained by a general negativity of “pandemic language,” as the analysis method is identical.
 
Indeed, this further supports our previous analysis of alternative news media generally painting a more negative picture, which goes hand-in-hand with an anti-systemic narrative that is typical for the majority of them \citep{Boberg_Quandt_Schatto-Eckrodt_Frischlich_2020}. The findings are also in sync with the notion that alternative news media used the crisis to portray things in a darker tone than did mainstream media. Conversely, from their perspective, the “legacy media” were not as negative, which may be linked to the assessment that these were less critical or even affirmative.
 
However, it needs to be stressed that the use of negative language is just one indicator, and even an alternative or critical viewpoint could also be expressed in positive language (e.g., ironically). Moreover, the data analysis is limited to a relatively small-time segment, without a baseline measurement from a “normal” period of coverage. So, these findings are just meant to be another puzzle piece in understanding the Corona coverage, and therefore, more nuanced research is needed (and indeed previewed for follow-up studies). 

\subsection{Fake news and conspiracy theories}\label{fake_news}

As in our previous analysis of alternative news media on the basis of the same data set \citep{Boberg_Quandt_Schatto-Eckrodt_Frischlich_2020}, we checked for the presence of fake news and conspiracy theories, as identified by fact-checking entities (RQ5). We applied the same definition here: “fake news” is totally or partially fabricated, with a made-up core of information \citep[for a discussion of the concept, see][]{Quandt_Frischlich_Boberg_Schatto-Eckrodt_2019}. In contrast to this, and following the definition by  \citet{Brotherton_2013}, conspiracy theories can contain truthful information, but they add so far unverified (and mostly unverifiable) claims, rumors, and embellishments, heterodox explanations, and a specific ideological reading of events (like malign intent and sensational subject matter). As discussed in our earlier research, fabricated news is easier to identify and debunk, while conspiracy theories are much harder to counter, as they are typically embedded into a larger, often total, belief system \citep{Schatto-Eckrodt_Boberg_Wintterlin_Quandt_2020}, which is self-insulating by principle \citep{Brotherton_2013}. Unsurprisingly, in the previous analysis of alternative news media’s Corona coverage, we found very limited use of fake news, but a more frequent reliance on conspiracy theories, and a recurring recourse on “deep state,” “evil elites,” and “oppression of the people” tropes, contributing to something that could be labeled as \emph{murky murmur}.  

Our current analysis of journalistic news media’s coverage on Facebook reveals a different focus (Table \ref{tab:fake_news}). The story on “Ibuprofen exacerbating COVID-19 according to a study from Vienna” was covered quite frequently here, while it rarely showed up on alternative news media. It is considered a fake news item by fact checkers, as the existence of the Vienna study was a fabrication. In contrast, there were just a few mentions of the conspiracy theories regarding a 2012 governmental report on the potential effects of an aggressive virus spread and the notion that Corona was intentionally designed in a lab, for example, as a bioweapon. Only the “Corona is not worse than the flu” theory gets some more mentions (but in relation to the respective overall output, this is still much lower than what we observed in alternative media). In that sense, the emphasis is reversed here when compared to the former analysis of alternative media, with more space in the coverage being devoted to the fake news item and less to the conspiracy theories. 

The emphasis on the Ibuprofen story may be surprising at first. Did the news media spread a high volume of a fake news item here? A closer inspection reveals a different picture, actually pointing in the opposite direction; indeed, many of the messages we found were directed against the fake news item. Further, some media covered the unclear situation regarding the claim when it first became public (noting the unverified nature of it), while others commented on an initial warning of the WHO against Ibuprofen (which was factual, in contrast to alleged study from Vienna, and most likely based on an unrelated note in The Lancet), and the later retraction of that warning by the WHO. Nearly all of the news items found here either informed about the story being fake news, the unverified and unclear information situation while the story was unfolding, or the parallel statements of the WHO and their retraction. 

We find a similar picture for the contextualization of conspiracy theories. As noted above, the two stories on the 2012 report and the ‘man-made virus’ were only mentioned a few times in posts of journalistic media on Facebook. Indeed, the latter one showed up on alternative news media channels about three times as often, despite the absolute number of posts being more than seven times lower, and it triggered more than 4,000 direct interactions there \citep{Boberg_Quandt_Schatto-Eckrodt_Frischlich_2020}, while we counted less than 200 interactions on mainstream journalistic media. Both of these stories are not reported in the form of conspiracy theories (in the sense of Brotherton’s definition), but either just referring to their true information core (in the case of the 2012 report) or in the context of stories on rumors and conspiracies circulating during the crisis (for the man-made virus theory).

\begin{landscape}
\begin{table}
 \caption{Overview of conspiracy theories and rumors}
 \centering
  \resizebox{!}{8cm}{
  \begin{tabularx}{\linewidth}{
  >{\raggedright\arraybackslash\hsize=.2\hsize}X%
  >{\raggedright\arraybackslash\hsize=.4\hsize}X%
  >{\raggedright\arraybackslash\hsize=.2\hsize}X%
  >{\raggedleft\arraybackslash\hsize=.08\hsize}X%
  >{\raggedleft\arraybackslash\hsize=.08\hsize}X%
  }
    \toprule
    Story & News media & Coverage & Posts & Interactions \\
    \midrule
Ibuprofen exacerbates progression of COVID-19 infection & Süddeutsche, Zeit online, Tagesspiegel, Hamburger Abendblatt, Sächsische.de, Ostsee-Zeitung & Unclear facts regarding Ibuprofen, negative effect not proven & 6 & 798 \\
\\[-0.8em]
 & Express, Welt, Focus, RP, Zeit online, Augsburger Allgemeine, Stern, Bild, Morgenpost, Braunschweiger Zeitung, B.Z., Nordbayern.de, Stuttgarter Nachrichten, taz, Neue Westfälische, HNA, Freie Presse, WAZ, FAZ, Donaukurier, Stuttgarter Zeitung, MAZ & Warning against fake news, Vienna University Hospital denies negative effect of Ibuprofen & 26 & 8666 \\
 \\[-0.8em]
 & Focus, Süddeutsche, PNP, Mitteldeutsche, Die Rheinpfalz, B.Z., Sächsische, SVZ, SWP, LVZ, NOZ, WN, Berliner Zeitung, Stuttgarter Zeitung, Heidelberg24, Tagesspiegel, WZ, HAZ, Freie Presse, tz München & WHO warns against potential negative effects of Ibuprofen & 22 & 8289 \\
 & InFranken.de & Experts (not further defined) warn against Ibuprofen & 1 & 527 \\
 \\[-0.8em]
 & FAZ, Stern, RP, SVZ, Mitteldeutsche Zeitung, Die Rheinpfalz, Stuttgarter Zeitung, HAZ, Ostseezeitung, Stuttgarter Nachrichten, MAZ, Hamburger Abendblatt & WHO withdraws warning about Ibuprofen & 12 & 2021 \\
 \\[-0.8em]
 & InFranken.de & Impending shortage of Paracetamol & 1 & 329 \\
\midrule
The German federal government already knew about the coming events as early as 2012 & Nordkurier, saechsische.de, Aachener Zeitung & Risk analysis of 2012 as a frightening scenario that might now occur or has already occurred & 3 & 381 \\
\midrule
Corona is a man-made laboratory virus & Passauer Neue Presse, FR, OVB & Mentioned in review of the most bizarre rumors and conspiracy theories related to Corona & 3 & 140 \\
\\[-0.8em]
& Wirtschaftswoche & Mentioned in review of the most bizarre rumors and conspiracy theories related to Corona, some of them with a true core &  1 & 35 \\
\midrule
Corona is not worse than a normal flu epidemic (Statements of Wolfgang Wodarg) & Spiegel, taz, WELT, Stern, Heilbronner Stimme, Berliner Zeitung, Tagesspiegel, B.Z., Express, KStA, Neue Westfälische, Schwäbische Zeitung, HNA, WZ, Watson, Cicero & Fact-checks and refutations of Wodarg’s statements & 17 & 11064 \\
\bottomrule
    \multicolumn{5}{l}{\emph{Notes:} \emph{N} = 18,051} \\
  \end{tabularx}
  }
  \label{tab:fake_news}
\end{table}
\end{landscape}

The only story related to conspiracy theories that triggered a notable number of messages, and a considerable number of interactions, was the “Corona is not worse than the flu” item, based on the statements of Wolfgang Wodarg. This 73-year-old politician and former physician publicly claimed the effects of COVID-19 for society to be minimal, and implied the whole crisis to be merely panic mongering. Further, alternative news media took his statements as a truthful oppositional voice, supporting the frequent claim that the whole pandemic was fabricated to serve a sinister plot and further strengthen the power of the elites. They also portrayed Wodarg as a relevant counter-opinion to one of the country’s leading virologists, Christian Drosten \citep{Boberg_Quandt_Schatto-Eckrodt_Frischlich_2020}. Again, journalistic media did not ignore the issue, but they addressed it in a different way; they did not report the ideas of Wodarg in the form of a (potentially true) conspiracy theory or rumor, but as statements that were currently triggering public reactions, especially on social media. In addition, they checked the basic facts behind his assumptions and refuted unfounded claims. It is interesting to see that this triggered a high number of interactions, which evidently mirrors the public interest in the debate on the potential threat of the virus. This public interest was plausibly the prime reason for journalistic media reporting on this case in the first place, as under other circumstances (and following typical relevance rules), the statements of a long-retired politician most likely wouldn’t have made headlines on multiple national media. 

Indeed, subsequently, journalistic media were also criticized by some for giving voice to Wodarg, while others criticized them for not offering even more oppositional voices (showing the contradictory expectations and viewpoints regarding journalistic coverage; see Section \ref{introduction}). This highlights the problems for the journalistic response, especially in an unprecedented crisis situation. It partially poses a catch-22 situation, as media need to report on unfolding topics of potential general importance and public interest, but they also generate attention, relevance, and focus through the reporting itself. If they report on a case like this, one can criticize them for over-pronouncing the opinions of a person who is neither in charge nor has proper expertise, just to serve the goal of impartiality \citep[a media bias known as “false balance”; see][]{Boykoff_Boykoff_2004}, but if they don’t do this, one can criticize them for neglecting a public debate. This situation is further complicated by the plausible assumption that journalists are aware of their power and responsibility in the process, making these decisions at least partially conscious and conflicting. This is not only an issue of simple gatekeeping and information selection, but an issue of constructing a media reality and determining in what ways this construction reflects the factual basis of a complex world plus the pertinence of the reported items for society.       

Summing up, while fake news and conspiracy theories were not absent from the Facebook messages of journalistic mainstream media, they were clearly marked as what they were, and often contextualized in meta-stories on fake news and conspiracy theories, rather than serving as the foundational information of the respective pieces. The question of whether media should have reported on these, or whether they should have ignored them for their irrelevance (in order to not step into the false balance trap), is one that cannot be answered on the basis of the current data alone, but it is certainly one that deserves further scientific attention.

\section{Discussion}\label{discussion}

The COVID-19 pandemic took both societies and news media around the globe by surprise. At first, the virus spread in China was deemed irrelevant for other countries, and reporting was insignificant. However, when the virus reached other countries, and both the worldwide and national impact became quickly apparent, many journalistic media switched to full crisis mode. The resulting explosion of concurrent, often complex and sometimes contradictory, information was quickly criticized both by public media experts and alternative voices on the Internet. Some observed signs of horse-race reporting and court circular, others were missing oppositional voices to official statements, some were unsatisfied with too much focus on Corona and a lack of plurality, while some claimed a lack of depth and background information—to name but a few of the claimed insufficiencies. However, these statements are based on individual and anecdotal observation. Empirical data on the performance of journalism during the crisis has been missing thus far.

The current study on the Facebook activities of German news media coverage during the early pandemic (beginning of January until second half of March), adds substantial empirical data to the debate. The analysis reveals the overall output of journalism being relatively constant during the early crisis, with a growth just during the last two analysis weeks when the Corona-related coverage began to dominate the news flow. Messages on the virus then upstaged all the other news. 

The overall topical structure of Corona-related news is varied and not uniform over time. Our analysis revealed various stages of the coverage, from early neglect of the topic in virtually all media, very limited response to the then contained initial infection cluster, to a rapid upscaling of activities when the uncontrolled infection wave began in late February. Then, we can see a massive growth and broadening of the pandemic coverage, leading to a multi-perspective view on virtually all societal fields. There are limited signs of horse-race reporting, especially during the very first reactions, but in contrast to public criticism, this type of news is not dominant. From the topical patterns we observed during the pandemic, we could generalize a phase logic of how journalism deals with the crisis—a model that may be tested against future data derived from other crises (even if they may be much different from what has been called an unprecedented crisis). 

In line with previous research, we could observe some concentration on the political elite in the coverage. Such a focus is not surprising, especially considering the role of these actors for decision processes during the crisis. Interestingly though, when compared to the coverage of alternative news media as examined in our previous working paper, the spectrum of actors was wider, and the concentration on elite politicians was much lower. Also, in contrast to alternative news media, institutions and companies were highly present as relevant societal group actors in journalistic mainstream media. In short, we didn’t find much support for an overwhelming amount of highly-targeted personalization in the current analysis, and only limited signs of a heroization of medical experts during the crisis (with only one virologist showing up in the top list of actors, which is actually the same person who appeared in the top list of alternative news media).   

The co-occurrence analysis of the main actors further revealed a very pronounced contextualization of the actors either in relation to specific events or representing specific fields of action. In particular, the main political actors were portrayed in relation to clearly defined roles and tasks in the crisis while some institutions and organizations were used to symbolize specific social issues in the coverage. This finding also hints at both the “information transfer” function of news in the portrayal of events, and the “world construction” function in explaining and (re)constructing the complexities of the social system that they observe and in which they operate.  

Indeed, the latter, (re)constructive function of news media has also been criticized in the crisis, as expert observers and alternative media alike noted a lack of distance to official institutions, politicians and “the elite,” resulting in an affirmative coverage and a one-sided portrayal of the situation, essentially constructing a biased or even false reality. While we did not analyze the intricacies of such world construction via the coverage, we checked the negativity in the analyzed messages, and found journalistic mainstream news media to be less negative in their coverage than what we found for alternative news media. This relative difference held true for both the Corona coverage and unrelated news. However, despite this generally less negative perspective of the journalistic mainstream, its pandemic news messages were still more negative than its other coverage. While this does not necessarily rule out the possibility of uncritical and affirmative crisis reporting, it also doesn’t support the notion of an overwhelming lack of negative evaluations in journalism during the crisis. 

Last but not least, we checked whether mainstream media’s news output contained fake news and conspiracy theories, as we found some to be present in the coverage of alternative news media in our previous analysis. Interestingly, when compared to alternative news media, conspiracy theories appeared less frequently, while one specific fake news story was covered more often. However, this does not mean that they spread such fabrications. On the contrary, conspiracy theories and fake news were consistently contextualized as what they are, and the coverage also included numerous posts debunking false claims circulating in public. 

Summing up, our analysis only mildly supports the criticism of journalism’s performance during the early crisis. On the one hand, there have been some signs of horse-race reporting and limitations of the coverage, especially in the initial stages of the emerging situation. Further, there were tendencies to primarily focus on elite actors, and journalism wasn’t overly negative in its reporting, as some critics have reported. On the other hand, some of these issues were much more pronounced in alternative news media according to our comparative analysis, which is somewhat ironic, given their permanent criticism of “the mainstream.” Also, some critics argued journalism to be panic mongering, which is neither supported by the negativity analysis nor the topical spectrum we found in our analysis. Grosso modo, we could not observe overwhelming signs of sprawling, but information-poor, pandemic news. So, in essence, there were no apparent systemic dysfunctionalities, although some aspects may deserve an improved handling in the future. 

As the data does not point at a radical failure of the system, we suspect that much of the stark criticism of journalism’s performance (as expressed, for example, in alternative media) is seemingly based on a very specific expectation, that is, of journalism to always be in basic opposition to politics or even more generally, the “system.” Indeed, that did not happen during the early crisis—at least there are no signs of a confrontational, negative, and anti-systemic coverage, which would have shown up in some of our analyses as well. However, such an expectation is based on a very narrow understanding of “critical” coverage to always be a knee-jerk rejection of “the other,” and to be against something. Instead, from our observation, journalism responded to the unprecedented situation differently—with a multi-perspective coverage of the crisis operating within the given societal system (again, as opposed to what we observed for alternative news media in our parallel analysis). 

It needs to be noted that our initial analysis of journalism’s Facebook posts in the Corona crisis is limited in several ways. We focused on one country (Germany), a specific segment of the pandemic development (the first three months), and a specific news distribution channel (Facebook). Further, we only applied a limited set of analyses, for example, excluding more intricate analyses of language and framing, and we refrained from hypothesis testing. More and refined research is needed to check for some of the assumptions we developed on the basis of this initial study. Most evidently, a further extension of the time period will reveal whether the news composition changes again, and whether these changes also lead to a more confrontational and conflicting coverage (something that could be anecdotally observed in the weeks that followed the sample period). Also, a more detailed sentiment analysis—and potentially, an analysis of specific types of jargon (like military language)—may be insightful for the understanding of how journalistic framing in times of crisis works. 

Overall, the current analysis contributes to the ongoing research efforts in understanding a very special and intense situation. In that sense, we hope that some of this remains idiosyncratic and that this type of situation won’t be recurring. Still, we also hope that with the findings outlined above, both scholars and practitioners gain new insight in the more general performance of journalism in a profound and global crisis.

\clearpage

\bibliography{references}

\end{document}